\documentclass[12pt,aps,english]{revtex4}

\usepackage{graphicx}
\usepackage{soul}
\usepackage{bm}
\usepackage{dcolumn}
\usepackage[usenames, dvipsnames]{color}
\usepackage{epsfig}
\usepackage{amsmath}
\usepackage{color}

\begin{document}

\title{Tracer diffusion in a sea of polymers with binding zones: mobile vs frozen traps}
\author{Nairhita Samanta and Rajarshi Chakrabarti*}
\affiliation{Department of Chemistry, Indian Institute of Technology Bombay, Powai, Mumbai 400076, E-mail: rajarshi@chem.iitb.ac.in}
\date{\today}

\begin{abstract}

We use molecular dynamics simulations to investigate the tracer diffusion in a sea of polymers with specific binding zones for the tracer.  These binding zones act as traps. Our simulations show that the tracer can undergo normal yet non-Gaussian diffusion under certain
circumstances, e.g, when the polymers with traps are frozen in space and the volume fraction and the binding strength of the traps are moderate. In this case, as the tracer moves, it experiences a heterogeneous environment and exhibits confined continuous time random walk (CTRW) like motion resulting a non-Gaussian behavior. Also the long time dynamics becomes subdiffusive as the number or the binding strength of the traps increases. However, if the polymers are mobile then the tracer dynamics is Gaussian but could be normal or subdiffusive depending on the number and the binding strength of the traps.  In addition, with increasing binding strength and the number of the polymer traps,
the probability of the tracer being trapped increases. On the other hand, removing the binding zones does not result trapping, even at  comparatively high crowding. Our simulations also show that the trapping probability increases with
the increasing size of the tracer and for a bigger tracer with the frozen polymer background the dynamics is only weakly non-Gaussian but highly subdiffusive. Our observations are in the same spirit as found in many recent experiments on tracer diffusion in polymeric materials and questions the validity of Gaussian theory to describe diffusion in crowded environment in general.

\end{abstract}

\maketitle

\section{Introduction}

In biology, chemistry and physics it is not uncommon to encounter a situation where the motion of a tagged particle shows deviation from the normal diffusive behavior \cite{Einstein, Chandrasekhar}. In other words,the long time limit of the mean square displacement $\left(\left<\overline{\delta^{2}(\tau)}\right>\right)$ of the tagged particle, which could be a biomolecule \cite{fradin, verkman, chowdhurycell2007, Elcock, Bandyopadhyay, metzlerprl2011}, polymer \cite{schweizerjpcb, fazli, chakrabartiphysica2, vanderzande, metzlerlooping}, nanoparticle \cite{schweizer2014} or a colloid bead \cite{stuart, hujpclet, granickpnas,metzlernjp2013}, does not scale linearly with time, or more precisely with the time difference $\tau$, but scales as $\left<\overline{\delta^{2}(\tau)}\right>\sim\tau^{\beta}$, where the exponent $\beta<1$ \cite{sokolov2012, weiss}. In other words the dynamics is subdiffusive. This happens due to the nature of the environment around the tagged particle, more often which is viscoelastic, crowded. Another important yet less explored aspect is the nature of the tagged particle  displacement distribution, which is not always Gaussian as suggested by the recent experiments \cite{granickpnas,granicknatmat,granickacsnano,hujpclet}. These experiments confirm diffusive yet a non-Gaussian dynamics of the tagged particle. For instance, in a recent single particle tracking experiment,
 Wang \textit{et al.} \cite{granickpnas} found that the diffusion of a colloid bead on phospholipid bilayer tubes to be Fickian, whereas the distribution
of the displacement of the bead was non-Gaussian. In fact, the distribution for long displacement was observed to be exponential \cite{granickpnas}.
In another independent study, the dynamics of a polystyrene nano-particle in polyethylene peroxide solution has been found to be
 normal yet non-Gaussian \cite{hujpclet}. Deviation from Gaussianity can emerge for different reasons, a continuous
time random walk (CTRW) process in a confined state or diffusion in a heterogeneous environment can also result in non-Gaussian distribution
of the displacement. It has also been reported that in a complex environment, a tracer can even have two diffusivities, one slower
and one faster \cite{granickmacmol, holm2013, karmakar}. Presence of two diffusivities \cite{holm2013} or a distribution of diffussivities \cite{karmakar} can also lead to a process which is non-Gaussian.

It was Chubnysky and Slater \cite{slater} who first came up with the idea of ``diffusing diffusivity" to explain normal yet non-Gaussian diffusion.  Their model could reproduce the observation of Wang \textit{et al.}. In the long time the distribution eventually becomes Gaussian following the Central limit theorem. However, an analytically exact model was still lacking until recently. Jain and Sebastian \cite{sebastian2016,sebastian2016jpcb} used path integral techniques to show that a random diffusivity model can lead to a normal yet non-Gaussian process and in their model, steady state solution of the diffusion equation for the diffusivity resulted an exponential distribution of the diffusivity as predicted in Chubnysky and Slater \cite{slater} formalism. All of these theoretical studies actually dealt with the case of static or dynamics disorder in diffusivity \cite{zwanzig, zwanzigjcp,sebastian2006, reichmanjpcb, chakrabarti2010}. In a very recent study, Cherstvy \textit{et al.}
have performed computer simulations of Langevin equation with random diffusivity to compare with the analytical results \cite{Cherstvymetzler2016}. Their observations also confirmed random diffusivity model as the one which can lead to normal yet non-Gaussian distribution.

Molecular dynamics simulations on model systems have been quite useful to shed light on tracer diffusion in crowded environment. Recently,
Ghosh \textit{et al.} have studied tracer diffusion in a heterogeneously crowded environment \cite{metzler2016} and subsequently using a continuum lattice made of static
obstacles \cite{metzlerpccp}. There have been
molecular dynamics simulation on tracer diffusion in an environment where the crowders are mobile, such as free polymers \cite{holm2013, kosovan2015}, polymers grafted in cylindrical channel \cite{chakrabartipre2013} or
in a polymer network \cite{Fatemeh}.  In addition, Kwon \textit{et al.}
have also looked into the tracer diffusion in the presence of crowding by taking care of the hydrodynamic interactions \cite{yethiraj2014}.  Most of these simulations have focused on homogeneous distributions of crowders, in reality, such as inside a biological cell, the environment is not only crowded but
has a heterogeneous distributions of sticky and non-sticky obstacles. McGuffee \textit{et al.} in their famous work modelled the bacterial cytoplasm in full atomistic details to perform Brownian dynamics simulation for the most abundant proteins in E. coli \cite{Elcock}, Hasnain \textit{et al.} used a coarse-grained model for the same \cite{Bandyopadhyay}.

In addition to crowding, a random or periodic external field can also lead to a deviation from normal Brownian diffusion. This can arise when a particle move through a heterogeneous medium with fluctuating interactions or topology \cite{Egelhaafepjst,Egelhaafpccp}. In this context it should be mentioned recently a molecular dynamics simulation has been performed with an all particles different (APD) system where each particle interact with another with a different potential \cite{rabinjcp}.

In this paper we investigate the tracer diffusion in a heterogeneous medium consisting of a collection of polymers with binding zones. These polymers essentially have specific binding zones acting as traps for the tracer. Excluding the binding zones the rest of the polymers serve as the non-sticky obstacles for the tracer with no binding affinity.  Thus our model is a combination of sticky and non-sticky obstacles and they are connected along a polymeric chain. On the other hand, the crowding is a consequence of the inclusion of many such chains in the simulation box. Therefore we study the effect of crowding and varying interaction on the tracer diffusion simultaneously.  The tracer diffusion is investigated in two different conditions, in one case, the polymers are placed randomly and allowed to move during the simulation, thus mimicking a mobile yet crowded environment. In another case, after randomly placing the polymers in the simulation box they are frozen to ensure a static heterogeneous distribution of sticky and non-sticky obstacles around the tracer.
We find the diffusion process to become non-Gaussian when the polymers are frozen as then the tracer experiences a heterogeneous distribution of sticky and non-sticky obstacles and shows jiggling motion in a cage followed by cage to cage jumps. But the diffusion becomes subdiffusive when the population of the
polymers is increased which resulted efficient trapping and becomes even more subdiffusive when the binding affinity of the trapping zones increases. On the other hand diffusion becomes Gaussian when the size of the tracer is increased.  This switching over to Gaussian from non-Gaussian diffusion on increasing the tracer size is also observed in a recent experiment \cite{hujpclet} on tracer diffusion in polymer gel.

The paper is arranged as follows. In section II we we present the simulation details, in section III we discuss the calculation methods.
The results and discussions are given in section IV and we conclude the paper in section V.

\section{Simulation details}

We perform molecular dynamics simulations using ESPResSo \cite{holmespresso}, a freely available molecular dynamics package. In our simulations Lennard-Jones parameters are used as the
unit system, where $\sigma_0$ is chosen as the unit of length and the unit of energy is given by $\epsilon_0$. All the particles in the system have identical masses.
\noindent Each of the polymers in the system are self avoiding and consists of twenty monomers.
The monomers are connected via finite extensible nonlinear elastic (FENE) potential.

\begin{equation}
V_{FENE}=-\frac{k_f r^2_{max}}{2}log\left[1-\left(\frac{r}{r_{max}}\right)^2\right]
\label{eq:fene}
\end{equation}

\noindent Where $k_f$ is the force constant of the bonds connecting each two monomers which can achieve a maximum length of $r_{max}$.
For our simulation the values of the parameters are $k_f=7, r_{max}=2, N=20$.

   \begin{figure}
 \centering
     \includegraphics[width=.4\textwidth]{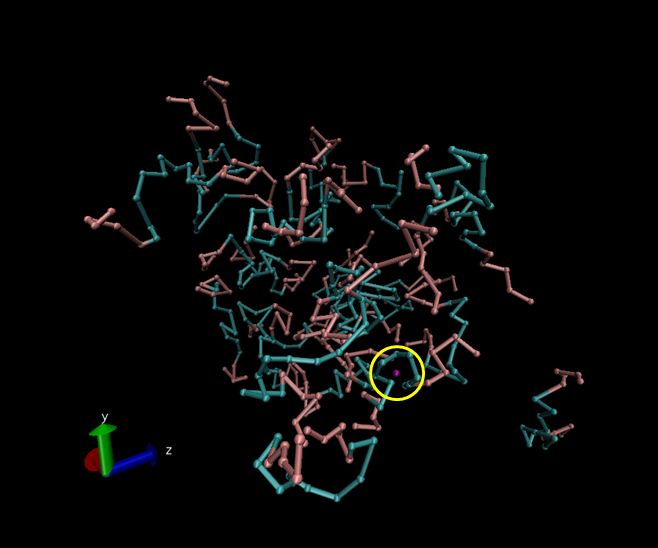}
   \caption A typical snap shot of the tracer and the polymers. The tracer is shown in purple (encircled by a yellow line for the convenience of the readers) and the binding zones of the polymers are shown in cyan.
   \label{fig:1}
 \end{figure}

\noindent The monomers and the tracers have same diameter of  $0.5\sigma_0$.
Among the twenty monomers in each polymer, ten monomers starting from sixth to fifteenth have attractive interaction with the tracer and
act as the binding zone as can be seen from Fig. (\ref{fig:1}) \cite{schultenvmd}. The binding zones interact with the tracer by means of the Lennard Jones (LJ) potential.

 \begin{equation}
V_{LJ}(r)=\begin{cases}4\epsilon\left[\left(\frac{\sigma}{r}\right)^{12}-\left(\frac{\sigma}{r}\right)^{6}\right], \mbox{if } r<r_{cut}\\
       =0, \mbox{otherwise}
\end{cases}
\label{eq:LJ}
\end{equation}

\noindent Here, $\sigma$ is the sum of radii of the two particles interacting via LJ potential and since all the particles have same size, here $\sigma=0.5\sigma_0$ and $\epsilon$ is varied from $2$ to $6$ with $r_{cut}=3\sigma_0$. We choose three particular values of $\epsilon$, $\epsilon=2\epsilon_0$, $\epsilon=4\epsilon_0$, $\epsilon=6\epsilon_0$.
Whereas, the rest of the monomers in the polymers are repulsive to the
tracer and this interaction is modelled by Weeks-Chandler-Anderson (WCA) potential \cite{andersen}.

\begin{equation}
V_{WCA}(r)=\begin{cases}4\epsilon\left[\left(\frac{\sigma}{r}\right)^{12}-\left(\frac{\sigma}{r}\right)^{6}\right]+\epsilon, \mbox{if } r<(2)^{1/6}\sigma\\
        =0, \mbox{otherwise}
\end{cases}
\label{eq:WCA}
\end{equation}

\noindent Where, $\sigma=0.5\sigma_0$, $r_{cut}=2^{1/6}\sigma_0$ and $\epsilon=1\epsilon_0$. The interaction between the monomers of the same and different polymers is always
repulsive and modelled by WCA potential with the same set of parameters mentioned above. This is just to ensure that these polymers do not form clusters. We investigate the tracer diffusion in different degrees of crowding.
For each value of $\epsilon$ in LJ potential the simulations are performed in three different monomer volume fractions $\phi$, namely $5\%$, $10\%$, $15\%$ which is achieved by changing the number of polymers in the system. However, the system remains in semi-dilute regime even at $\phi=15\%$. The values of parameters of WCA potential remain the same in every simulation.
For each set of $\epsilon$ and $\phi$ we generate thirty trajectories of the tracer. For each simulations the time step
($\delta t$) is chosen to be $0.001$ and after equilibrating the system long enough so that polymers have relaxed, the final simulation are
carried out for $25\times10^{4}$ steps. To accelerate the simulations we record the position of the tracer and each monomers in the system at every $50^{th}$ step. Therefore we obtain the data for total $5000$ steps at every $0.05$ time difference.
We use Langevin thermostat in NVT ensemble and use velocity Verlet algorithm for the integration of each time step.

\noindent The dynamics of each particle in the system is described by the Langevin equation

\begin{equation}
m\frac{d^2 r(t)}{dt^2}=-\xi\frac{dr}{dt}-\bigtriangledown\sum_{i} V(r-r_{i})+f(t)
\label{eq:langevin}
\end{equation}

\noindent Here, $m$ is the mass of the particles, $\xi$ is the friction coefficient which is considered to be $\xi=1$ always. $r(t)$ is the position of the particle at time $t$ and $f(t)$ is the random force acting on it. The random force $f(t)$ is a white noise with zero first moment \cite{doibook}.

\begin{equation}
   \left<f(t)\right>=0,
   \left<f_{\alpha}(t^{\prime})f_{\beta}(t^{\prime\prime})\right>=2 \xi k_B T \delta_{\alpha\beta}\delta(t^{\prime}-t^{\prime\prime})
\label{eq:random-forcerouse}
\end{equation}

\noindent Where, $k_B$ is the Boltzmann constant and $T$ is the temperature of the thermostat and the thermal energy, $k_BT=1$. As shown above all the particles experience Gaussian distributed white noise. The sum in Eq. (\ref{eq:langevin}) is over the position of all the particles in the system excluding the one being evaluated. We do not consider any hydrodynamic interaction in our simulations.

\section{Calculation methods}

To monitor the tracer diffusion we compute the Mean square displacement $\left(\left<\overline{\delta^{2}(\tau)}\right>\right)$ of the tracer. The time-averaged $MSD(\tau)$ is given by

\begin{equation}
\overline{\delta^{2}(\tau)}=\overline{[r(t+\tau)-r(t)]^2}
\label{eq:msd}
\end{equation}

\noindent Where $r(t+\tau)$ is the position of tracer at time $(t+\tau)$ and $r(t)$ is the same at the initial time $t$. The average is done over all the initial values ($t$). We also carry out ensemble average of the time-averaged $\left<\overline{\delta^{2}(\tau)}\right>$ over all thirty different trajectories for the tracer. For Fickian diffusion after the initial ballistic region the $\left<\overline{\delta^{2}(\tau)}\right>$ is linearly proportional to the time difference i.e. $\left<\overline{\delta^{2}(\tau)}\right>\sim\tau^\beta$, where, $\beta=1$. Whereas, for a subdiffusive process $\beta<1$. \\

\noindent To probe the nature of dynamics further, we calculate the velocity autocorrelation function ($C_v(\tau)$)

\begin{equation}
C_v(\tau)=\left<\vec{v}(t+\tau).\vec{v}(t)\right>/\left<v^2(t)\right>
\label{eq:cv}
\end{equation} \\

\noindent For normal Brownian motion $C_v(\tau)$ is exponential whereas negative correlation at short $\tau$ can originate from either fractional Brownian motion or Continuous time random walk (CTRW) in the presence of confinement \cite{sokolov2012, sain2013}. In the long time it approaches zero. \\

\noindent  Now to probe whether the tracer diffusion is Gaussian or not, we chose to calculate the non-Gaussianity parameter ($\alpha_2(\tau)$). The non-Gaussianity parameter is used extensively in the literature especially in connection to glassy dynamics \cite{schweizer2006, Zamponi}. However we do not have any glass like behavior here as the volume fraction of the polymers are below the onset of glass transition. For a three dimensional process $\alpha_2(\tau)$ is given by

\begin{equation}
 \alpha_2(\tau)=\frac{3\left<\delta r^4(\tau)\right>}{5\left<\delta r^2(\tau)\right>^2}-1
\label{eq:ngp}
\end{equation}

\noindent One can easily check that the non-Gaussianity parameter is exactly zero for a free diffusion with Gaussian distribution. Whereas, for non-Gaussian process e.g. CTRW will show a deviation from zero \cite{metzlerpccp, weiss, schweizer2006, michels}. CTRW arises when a tracer occasionally stops at intervals and as a result has a long tailed distribution of waiting times \cite{sain2013}.

\section{Results and discussions}

In this section we discuss and analyze the simulation results. For analysis we relied on our codes. Whether the tracer dynamics is diffusive or subdiffusive is interpreted by analyzing the mean square displacement $\left(\left<\overline{\delta^{2}(\tau)}\right>\right)$ of
the tracer. As already mentioned we consider two different cases of the the tracer dynamics,
one where the polymers are initially randomly placed and allowed to move during the simulation and another case where the polymers are frozen after placing them randomly. Thus we have two cases, in one the tracer diffuses
in presence of randomly placed but static obstacles and in the other case these obstacles are mobile. We investigate the dynamics of the tracer at varying volume fractions ($\phi$) by changing the number of polymers in the simulation box and also carry out the simulations for a range of binding strengths ($\epsilon$) between the tracer and the polymer traps. For simplicity we do not write $\sigma_0$ and $\epsilon_0$ in the rest of the paper. For example, $\epsilon=2\epsilon_0$ is written as $\epsilon=2$.

\subsection{Mean Square Displacement $\left<\overline{\delta^{2}(\tau)}\right>$}

\begin{figure}
  \centering
     \begin{tabular}{@{}cccc@{}}
       \includegraphics[width=.5\textwidth]{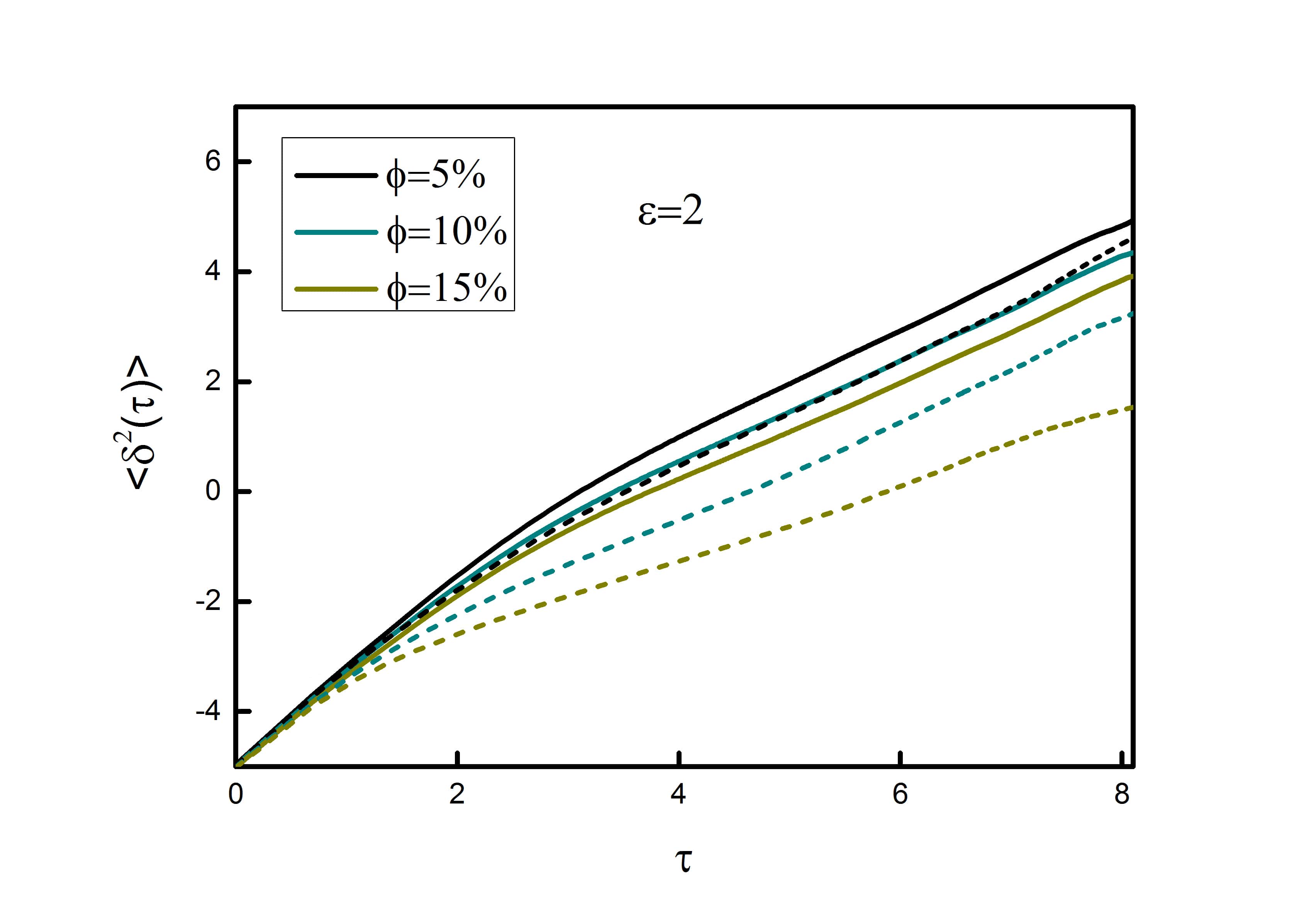} (a)
       \includegraphics[width=.5\textwidth]{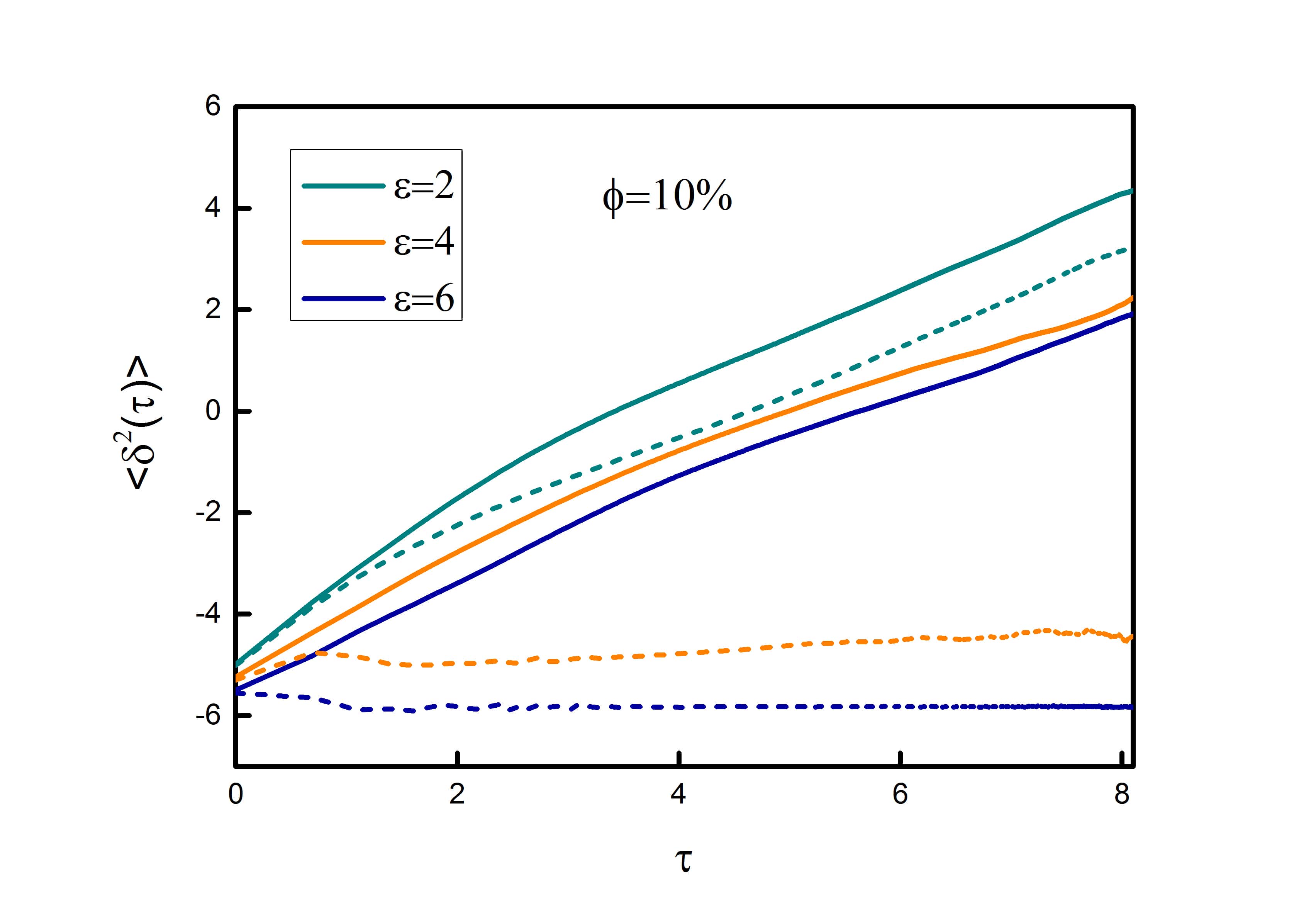} (b)
         \end{tabular}
     \caption{Log-log plot of the $\left<\overline{\delta^{2}(\tau)}\right>$ vs $\tau$ (a) at different volume fractions ($\phi$) (b) at different binding affinities ($\epsilon$). The
solid lines correspond to the diffusion in the presence of mobile polymers, dashed lines correspond to frozen polymers.}
     \label{fig:2}
   \end{figure}

\begin{figure}
   \centering
     \begin{tabular}{@{}cccc@{}}
       \includegraphics[width=.5\textwidth]{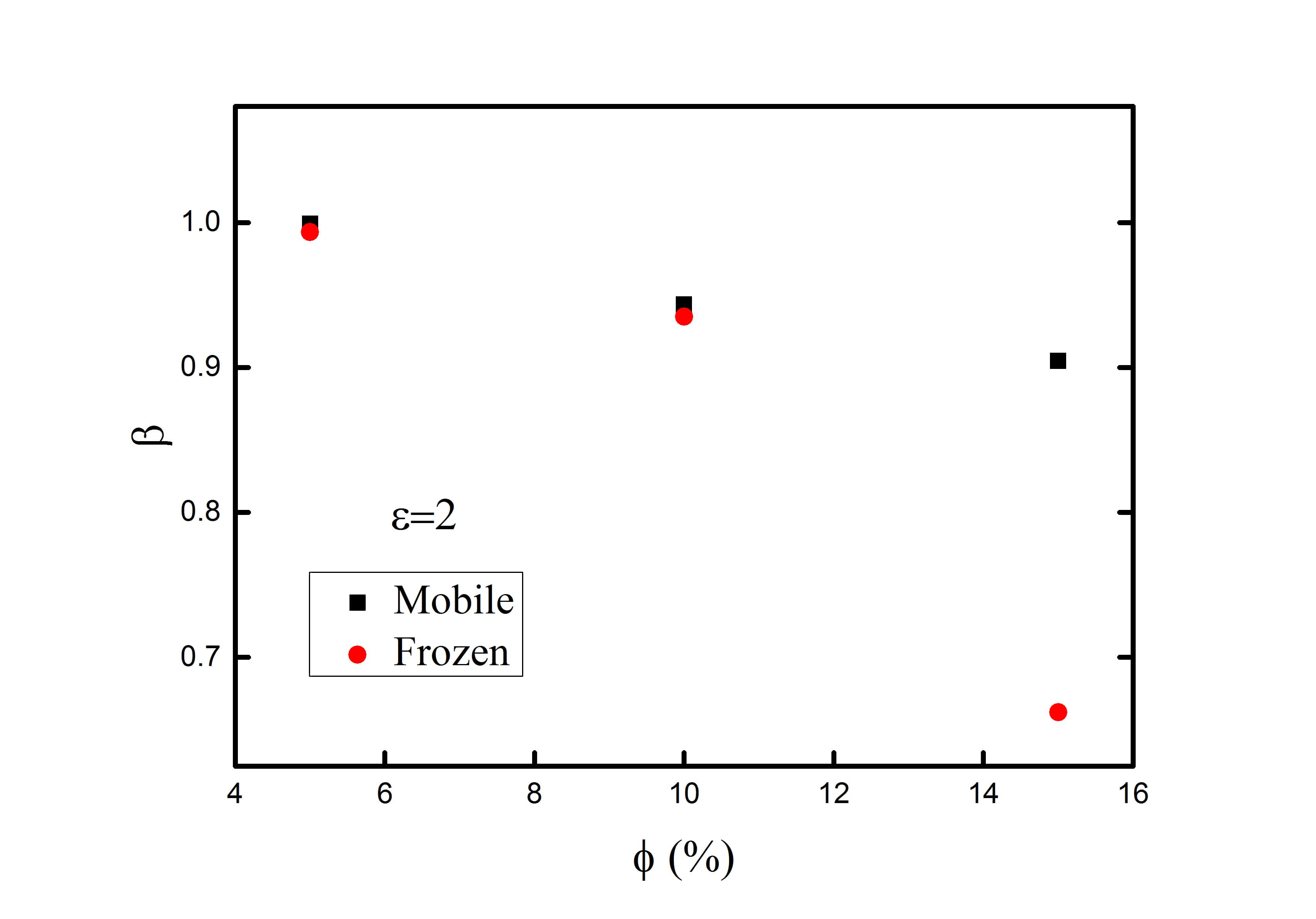} (a)
       \includegraphics[width=.5\textwidth]{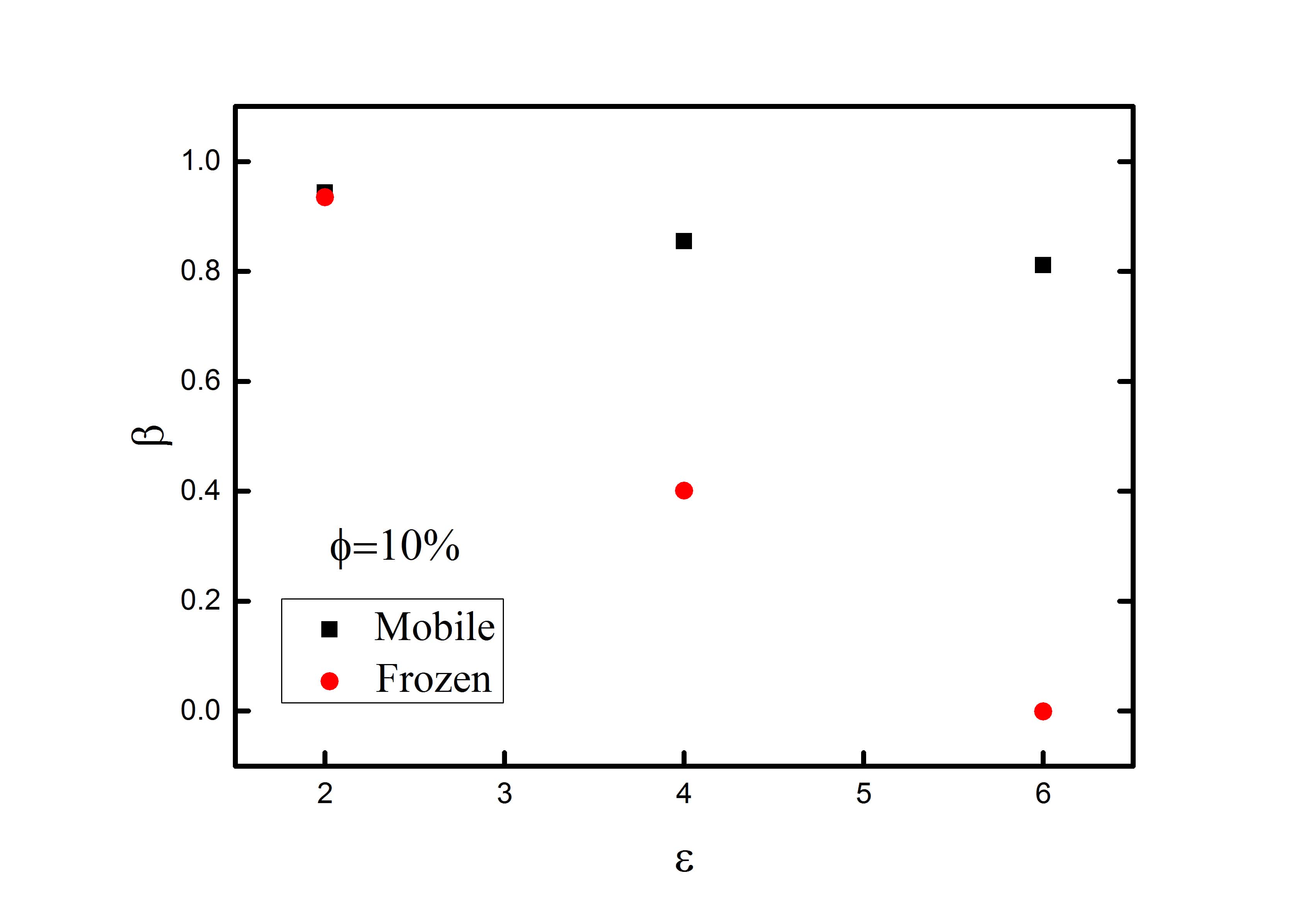} (b)
         \end{tabular}
     \caption{Plots of diffusion exponents (a) against the volume fraction ($\phi$) (b) against the binding affinity of the trapping zones ($\epsilon$).}
     \label{fig:3}
       \end{figure}

In Fig.(\ref{fig:2}a) we show the log-log plot of time-ensemble averaged $\left<\overline{\delta^{2}(\tau)}\right>$ (ensemble average of Eq. (\ref{eq:msd})) against the time difference ($\tau$) at different
volume fractions ($\phi$) with the binding affinity ($\epsilon$) of the trapping zones of the polymers fixed at $\epsilon=2$. The solid lines correspond to the $\left<\overline{\delta^{2}(\tau)}\right>$ of the tracer in the presence
of mobile polymers while the dashed lines refer to the same in the presence of frozen polymers. The $\left<\overline{\delta^{2}(\tau)}\right>$ of the tracer passes through a ballistic regime
at short time.
As the volume fractions increases, the $\left<\overline{\delta^{2}(\tau)}\right>$ of the tracer grows slowly. In the presence of frozen polymers, $\left<\overline{\delta^{2}(\tau)}\right>$ grows even slower
in comparison to the same in the presence of mobile polymers. This happens since, in case of mobile polymers the movement of the polymers
facilitate the movement of the tracer as well. Whereas, in case of frozen polymers if the tracer once gets trapped in the binding zones of the polymers,
 the probability of being trapped for longer time is higher
in the absence of any fluctuations coming from the polymers. In Fig. (\ref{fig:2}b) the  $\left<\overline{\delta^{2}(\tau)}\right>$ is shown for a range of binding strengths of the trapping
zones at a fixed volume
fraction, $\phi=10\%$. Here also we compare the effect of mobile as well the frozen polymers on the tracer. Due to the higher binding affinity with increasing value of $\epsilon$,
the tracer tends to bind to the trapping zones for longer time and this results in the lower $\left<\overline{\delta^{2}(\tau)}\right>$s. Although in the presence of mobile polymers
the tracer starts with
the ballistic motion and crosses over to Brownian motion, in the presence of frozen polymers at very high $\epsilon$ the $\left<\overline{\delta^{2}(\tau)}\right>$
practically remains unchanged. Next we calculate the diffusion exponents ($\beta$) from the plot of $\left<\overline{\delta^{2}(\tau)}\right>$ vs $\tau$, in the long time limit $\left<\overline{\delta^{2}(\tau)}\right>\sim\tau^{\beta}$.
Fig. (\ref{fig:3}) shows diffusion exponents of the tracer at different volume fractions and binding affinities. From Fig. (\ref{fig:3}a) it is observed that the tracer undergoes normal Brownian motion when the volume fraction is low ($\phi=5\%$).
As the volume occupancy by the polymers increases the tracer becomes subdiffusive and value of $\beta$ drops below $1$. In the presence of mobile polymers at
higher volume fraction the tracer is very weakly subdiffusive
whereas in the presence of immobile polymers the tracer exhibits strong subdiffusion. As shown in Fig. (\ref{fig:3}b),
 the tracer shows very strong subdiffusion when the binding affinities of the trapping zones are increased keeping the volume fraction constant. In fact, in
 the presence of frozen polymers when the value of $\epsilon=6$, $\beta$ is negligible confirming that the tracer hardly moves and the dynamics becomes non-ergodic. This is consistent with the $\left<\overline{\delta^{2}(\tau)}\right>$ plot and it implies
that the tracer remains trapped for most of the simulation time when the polymers are immobile and binding strength of the trapping zones are high. However, if the simulations are run for a very long time the diffusions should cross over to Brownian motion in every case. From the two set of our chosen values of $\epsilon$ and $\phi$, we see the effect of higher binding
 affinity on diffusion precess is more profound in comparison to the volume occupancy by the polymers.

\subsection{Velocity Autocorrelation ($C_v(\tau)$)}

\begin{figure}
  \centering
    \begin{tabular}{@{}cccc@{}}
      \includegraphics[width=.5\textwidth]{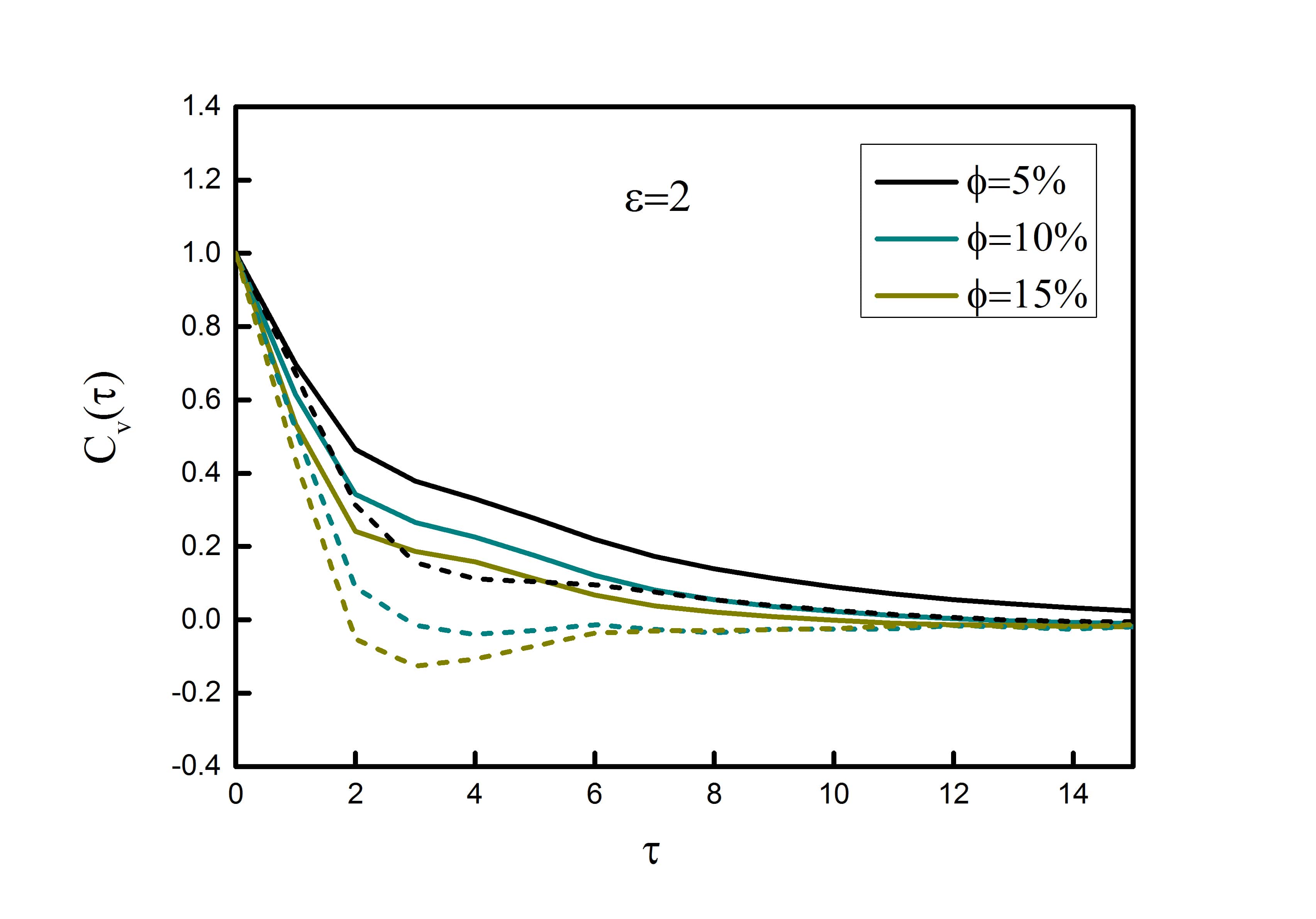} (a)
      \includegraphics[width=.5\textwidth]{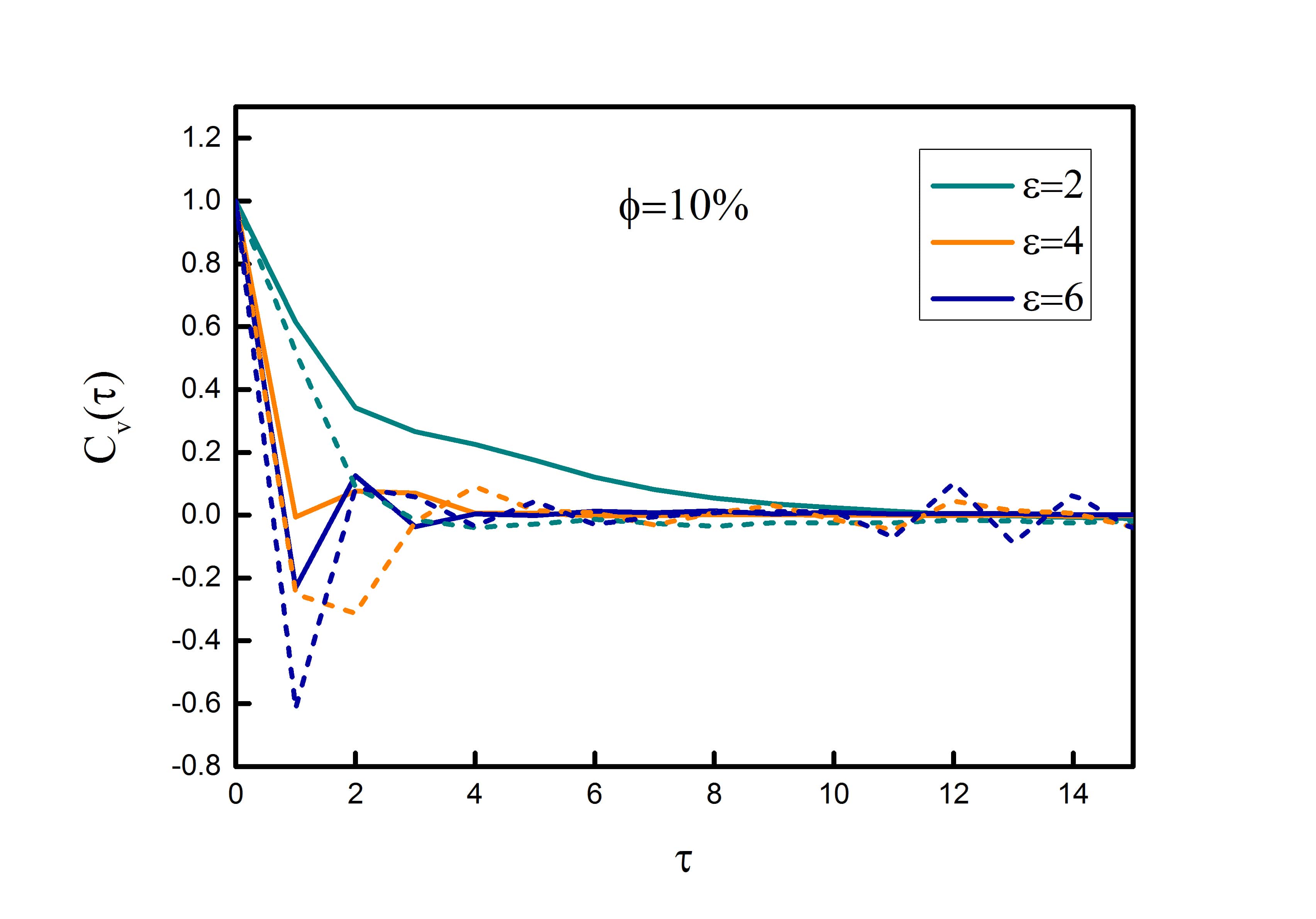} (b)
        \end{tabular}
    \caption{Log-linear plot of the velocity autocorrelation function ($C_v(\tau)$) against time ($\tau$) (a) at different volume fractions, (b) at different binding strengths between the tracer and the polymer traps. The
solid lines correspond to the diffusion in the presence of mobile polymers, dashed lines correspond to frozen polymers.}
    \label{fig:4}
  \end{figure}

\begin{figure}
 \centering
     \includegraphics[width=.6\textwidth]{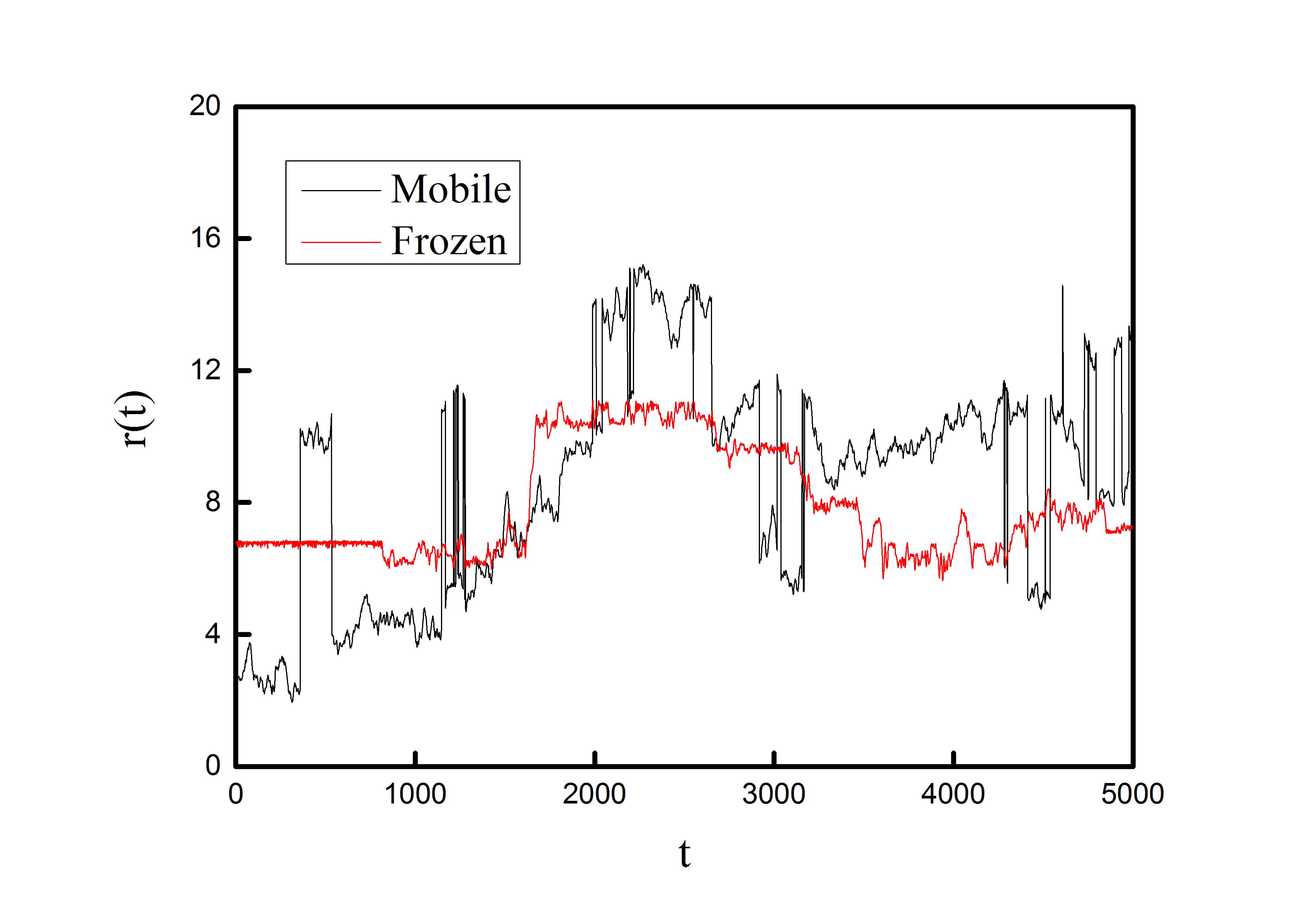}

   \caption{Trajectories of the tracer (in absolute time) in the presence of mobile and frozen polymers at $\phi=15\%$ and $\epsilon=2$. On freezing the polymers, cages become stable, while with mobile polymers cages are transient.}
   \label{fig:11}
 \end{figure}

There can be different types of subdiffusive processes \cite{metzlerpccp2014} and the nature of the dynamics can be further confirmed by analyzing the velocity autocorrelation ($C_v(\tau)$) of the tracer as defined in Eq. (\ref{eq:cv}). $C_v(\tau)$ vs $\tau$ is shown in Fig. (\ref{fig:4}). In the presence of mobile polymers at $\epsilon=2$, $C_v(\tau)$ is always positive and the trend of the correlation loss is consistent, and
Fig. (\ref{fig:4}a) shows with increasing volume fraction, correlation decay is faster. However, when the polymers are frozen in space,
at higher volume fraction, the correlation becomes negative at short time. With increasing binding affinity even more pronounced negative correlation is observed as can be seen from
Fig. (\ref{fig:4}b).
For mobile polymers, only higher $\epsilon$ gives rise to negative auto-correlation.
Such negative correlation can emerge primarily from two different mechanisms, first is fractional Brownian motion \cite{sain2013} and the second is confined CTRW  \cite{weitz, sain2013, spakowitzbioj}. Emergence of such negative correlations with frozen polymers confirms confined CTRW type motion. This can be also seen from the trajectory (Fig. (\ref{fig:11})) which shows motion within a cage formed by the polymer chains, followed by a big jump to another cage. In this case these cages are in the order of $\sim 2$ times the tracer size and static, since the polymers are frozen. Within the cage the tracer jiggles around and frequently changes the direction of its motion, contributing to the negative part of $C_v(\tau)$. While with mobile polymers cages are hardly formed as these polymers do not stick to each other and if formed these are only transiently stable (Fig. (\ref{fig:11})). This explains why only very weak negative correlations in $C_v(\tau)$ are seen with mobile polymers and that is also only at high values of $\phi$ and $\epsilon$. However, these negative correlations can also arise due to the viscoelasticity of polymers.

\subsection{Non-Gaussianity Parameter ($\alpha_2(\tau)$)}

\begin{figure}
 \centering
   \begin{tabular}{@{}cccc@{}}
     \includegraphics[width=.5\textwidth]{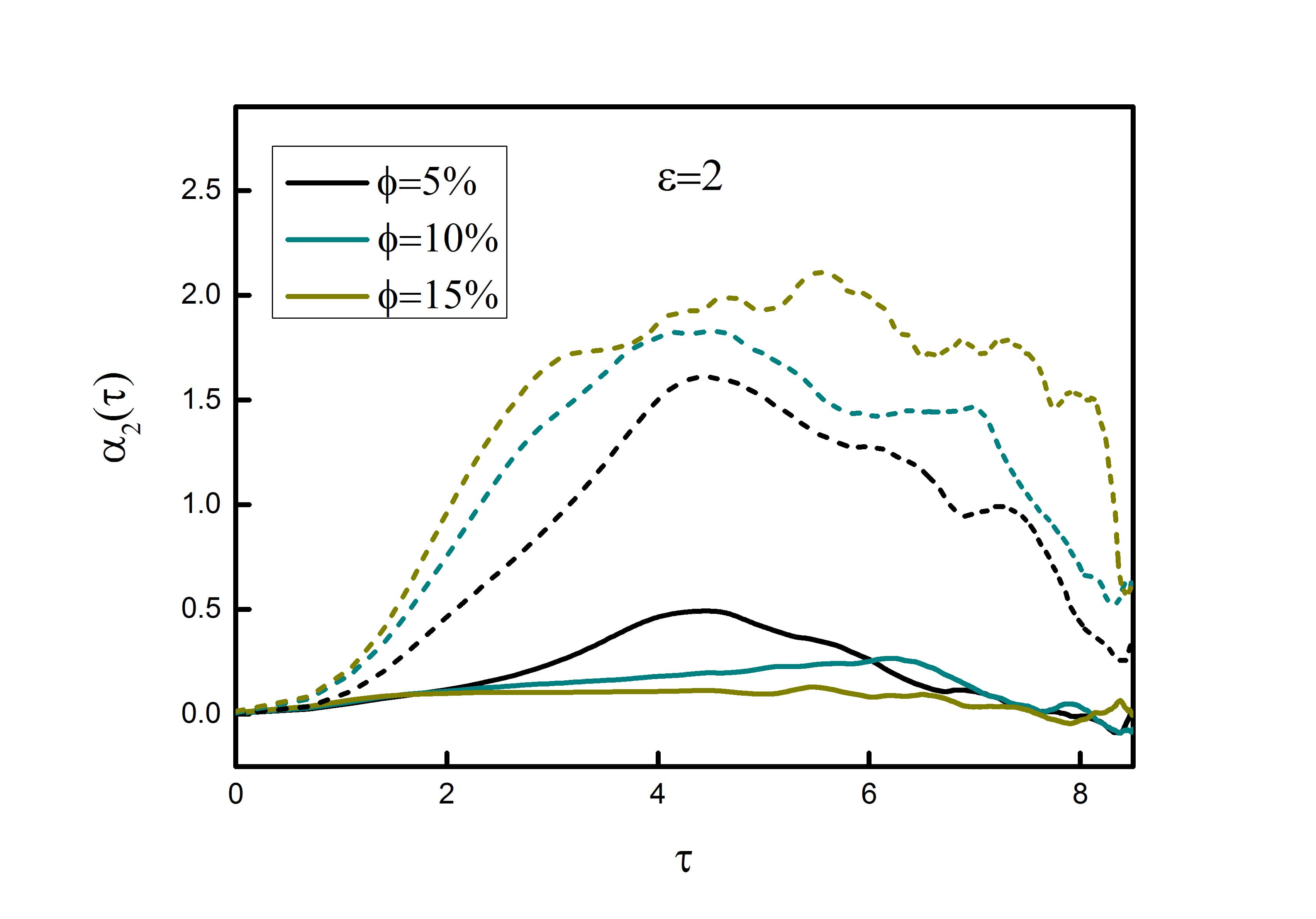} (a)
     \includegraphics[width=.5\textwidth]{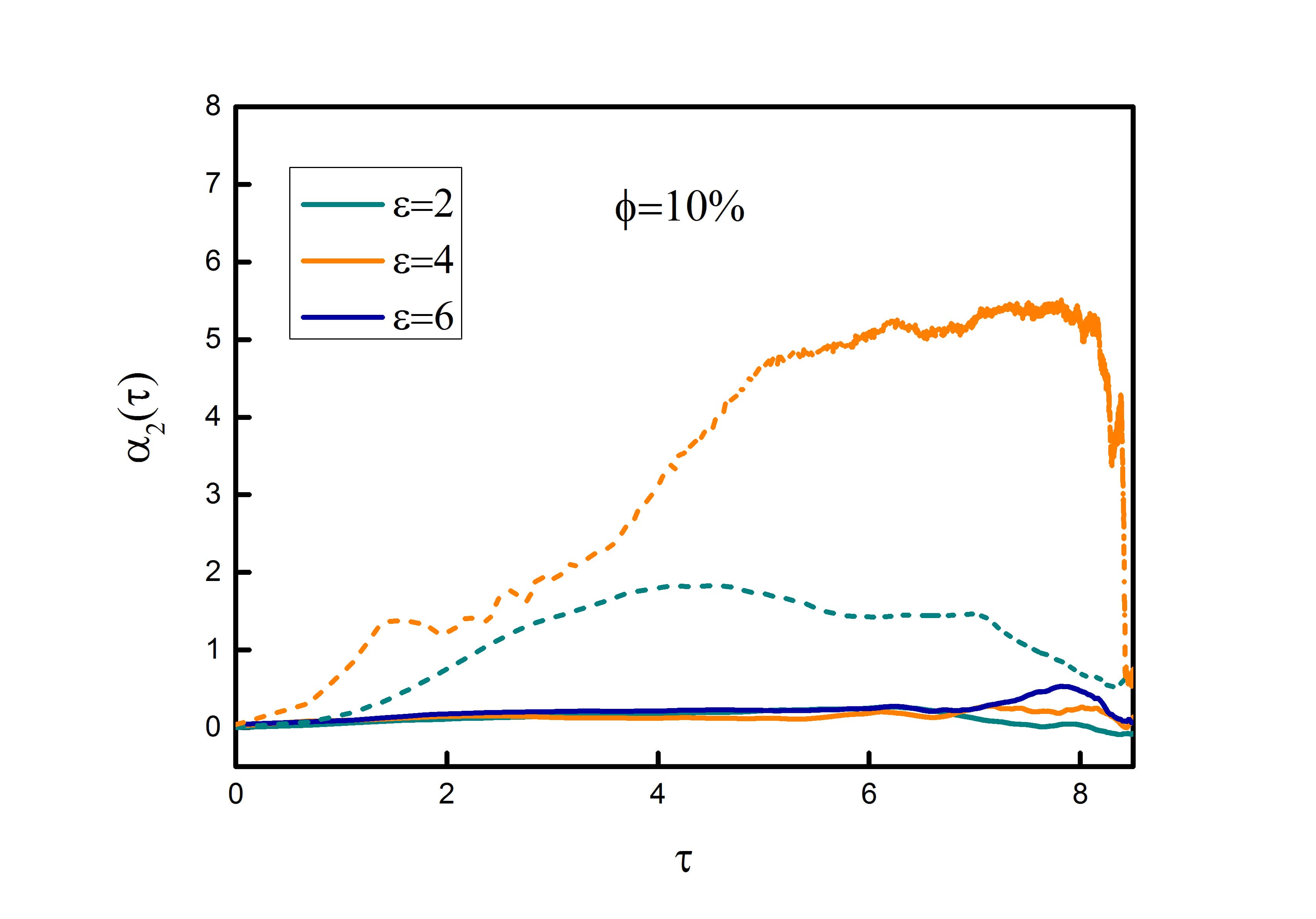} (b)
       \end{tabular}
   \caption{Log-linear plot of the non-Gaussianity parameter ($\alpha_2(\tau)$) against time ($\tau$) (a) at different volume fractions, (b) at different binding strengths between the tracer and the polymer traps. The
solid lines correspond to the diffusion in the presence of mobile polymers, dashed lines correspond to frozen polymers.}
   \label{fig:5}
 \end{figure}

 To probe the tracer dynamics deeper,  we also calculate the non-Gaussianity parameter, $\alpha_2(\tau)$ defined in Eq. (\ref{eq:ngp}) . It is known that any distribution apart from Gaussian gives rise to non-zero
$\alpha_2(\tau)$. Fig. (\ref{fig:5}a) shows $\alpha_2(\tau)$ deviates very slightly from zero when the tracer diffuses in the presence of mobile polymers. This implies initially the diffusion is only very weakly non-Gaussian. This is presumably due to the fact that on an average the tracer sees a crowded yet homogeneous environment. This is further established by a vanishing $\alpha_2(\tau)$ at long $\tau$. Whereas, in case of
frozen polymers, pronounced deviation can be noticed as observed by Saltzman and Schweizer in glassy hard sphere fluids  \cite{schweizer2006}, the maximum values of the plots of $\alpha_2(\tau)$ vs time
increases with increasing volume fraction. Eventually at long time all the processes become Gaussian.
From the values of the diffusion exponent, it is already observed that when the volume fraction is low the tracer undergoes normal Brownian diffusion. However from the values of
non-Gaussianity parameter it can be seen even when the volume fraction is low, the distribution of displacement is not Gaussian for the tracer. Although the deviation is small in the case of mobile polymers, it shows strong non-Gaussian behavior in the presence of immobile polymers. This trend is similar as observed in the some recent experiments
\cite{granickpnas, granickacsnano, hujpclet}. But at higher volume fraction the diffusion is anomalous and non-Gaussian.
In the presence of frozen polymers, the deviation from Gaussianity can emerge from confined CTRW process. This is also validated from the negative velocity auto-correlation
observed in this case. Fig. (\ref{fig:5}b) shows the plots $\alpha_2(\tau)$ at different values of $\epsilon$. In case of mobile polymers again very weak deviation is observed.
While with frozen polymers, the deviation from zero increases with increasing $\epsilon$ and
the magnitude of the deviation for $\epsilon=4$ is very high which lasts for very long time as well. However, at $\epsilon=6$, $\alpha_2(\tau)$ shows
almost no deviation (not shown). This might seem very surprising at first, but from the  $\left<\overline{\delta^{2}(\tau)}\right>$ and the diffusion exponent $\beta$ it is clearly observed that the tracer shows almost no movement when the binding affinity of the frozen polymers are very high and that gets reflected in the almost negligible value of the non-Gaussianity parameter.

\subsection{Trapping time}

\begin{figure}
 \centering
   \begin{tabular}{@{}cccc@{}}
     \includegraphics[width=.5\textwidth]{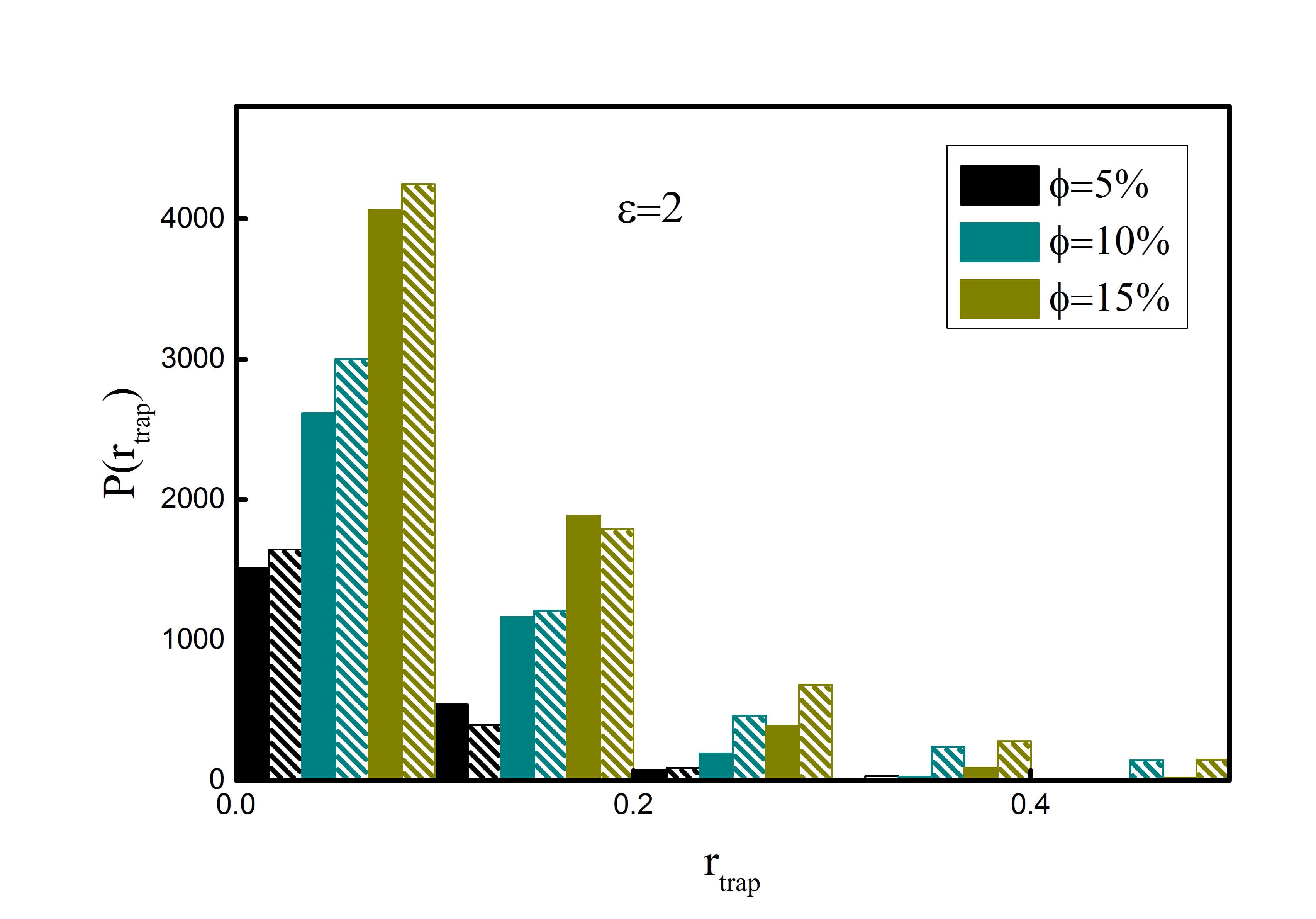} (a)
     \includegraphics[width=.5\textwidth]{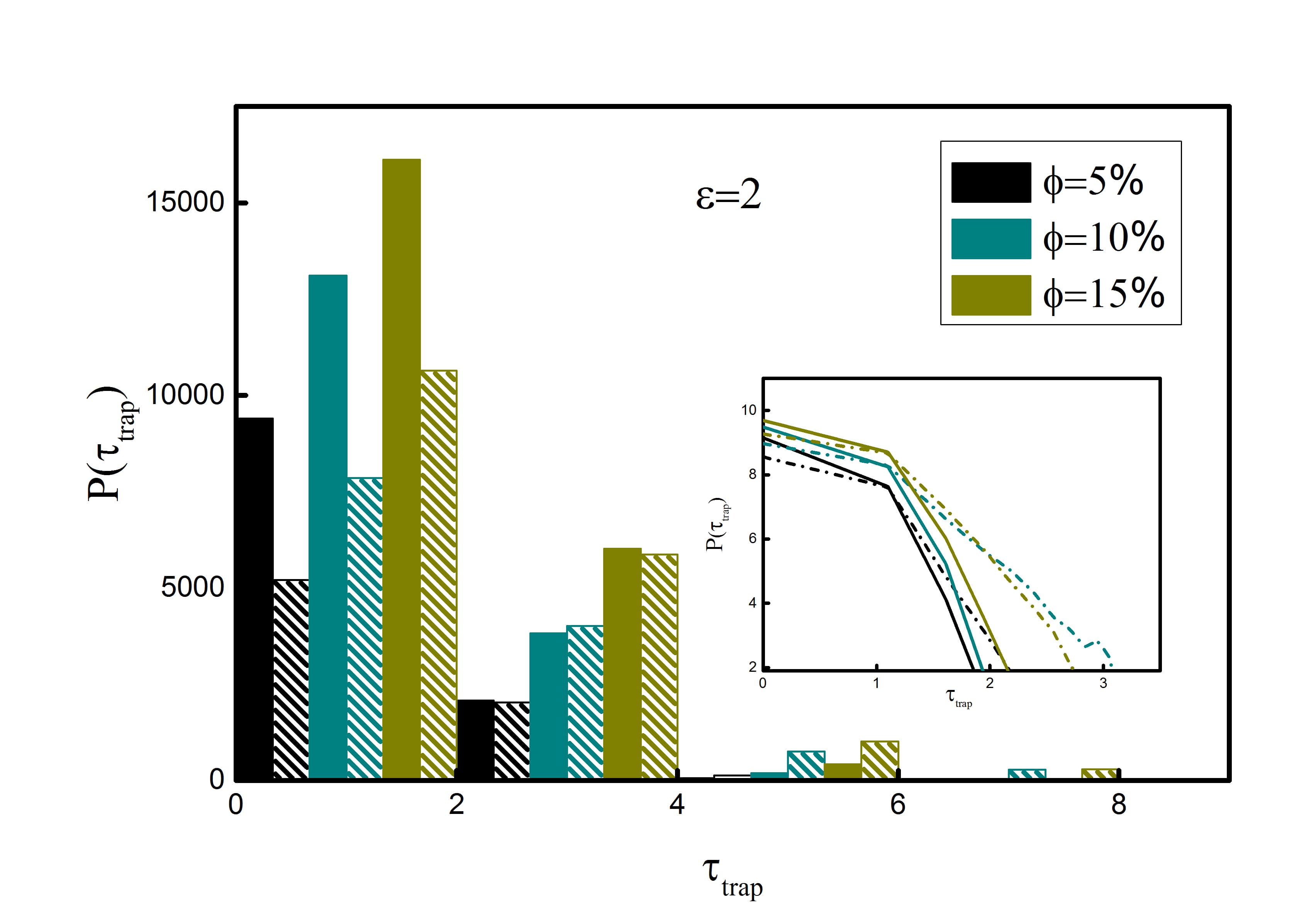} (b)
       \end{tabular}
   \caption{Histograms of (a) the distance travelled by the tracer in the trapped state (b) the time spent in the trapped state at different degrees of volume occupancy. The inset shows the log-log plot of the distribution of the trapping time.} The solid bars correspond to the tracer in the presence of mobile polymers and the bars filled with dashed lines correspond to the tracer in the presence of frozen polymers.
   \label{fig:6}
 \end{figure}

\begin{figure}
 \centering
   \begin{tabular}{@{}cccc@{}}
     \includegraphics[width=.5\textwidth]{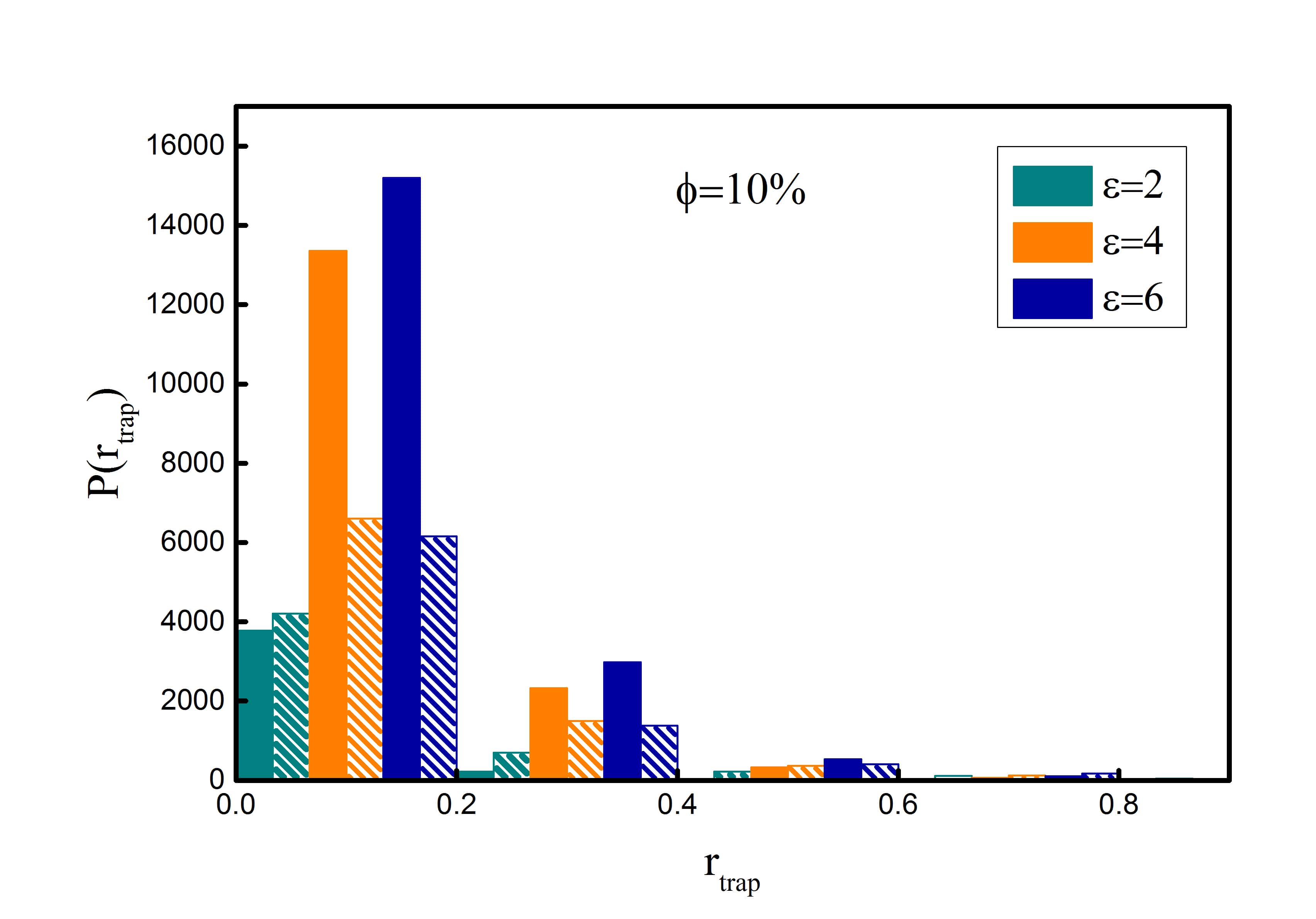} (a)
     \includegraphics[width=.5\textwidth]{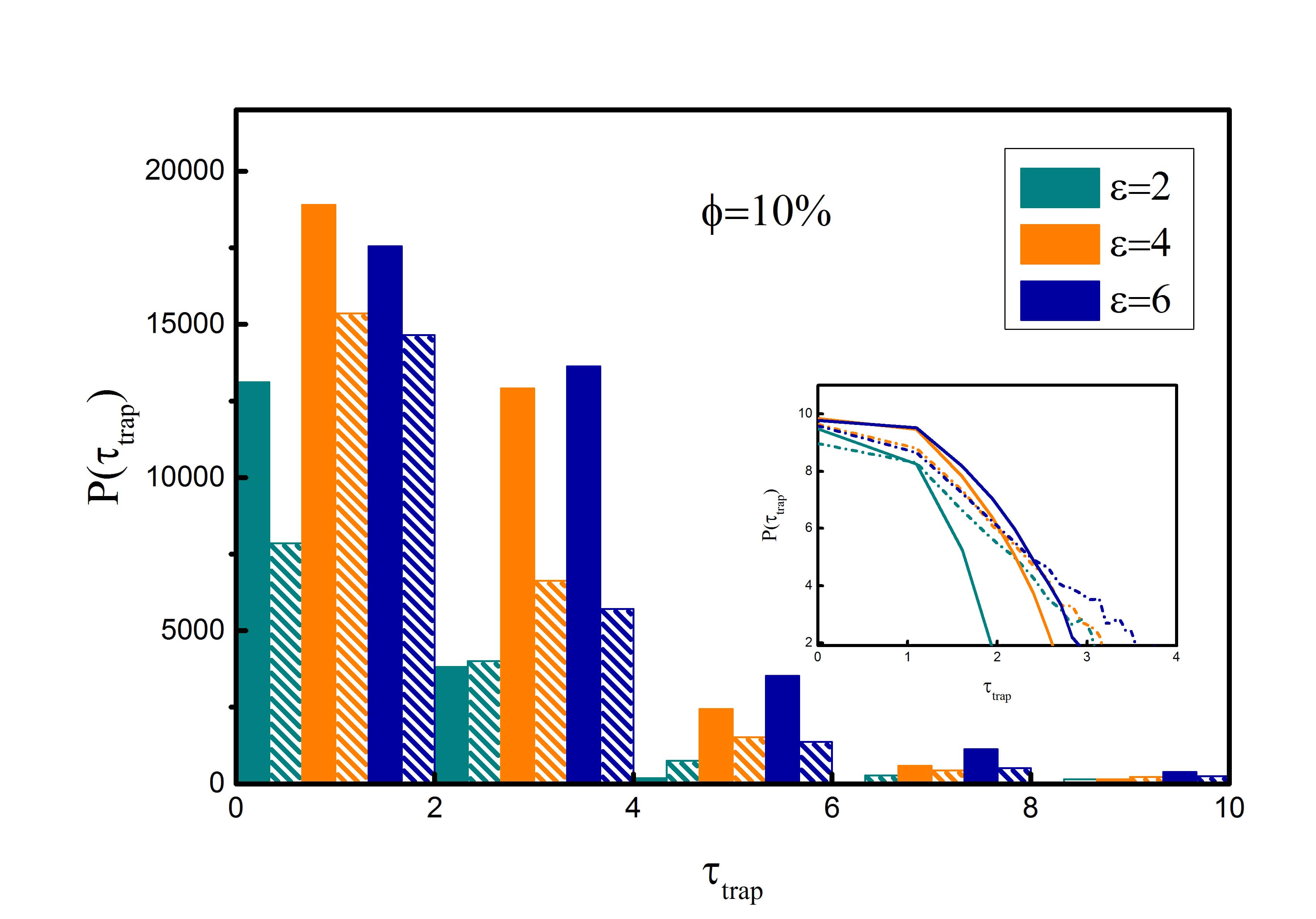} (b)
 \end{tabular}
   \caption{Histograms of (a) the distance travelled by the tracer in the trapped state (b) the time spent in the trapped state at different levels of tracer binding zone interaction. The inset shows the log-log plot of the distribution of the trapping time.} The solid bars correspond to the tracer in the presence of mobile polymers and the bars filled with dashed lines correspond to the tracer in the presence of frozen polymers.
   \label{fig:7}
 \end{figure}

\begin{figure} \centering
   \begin{tabular}{@{}cccc@{}}
     \includegraphics[width=.5\textwidth]{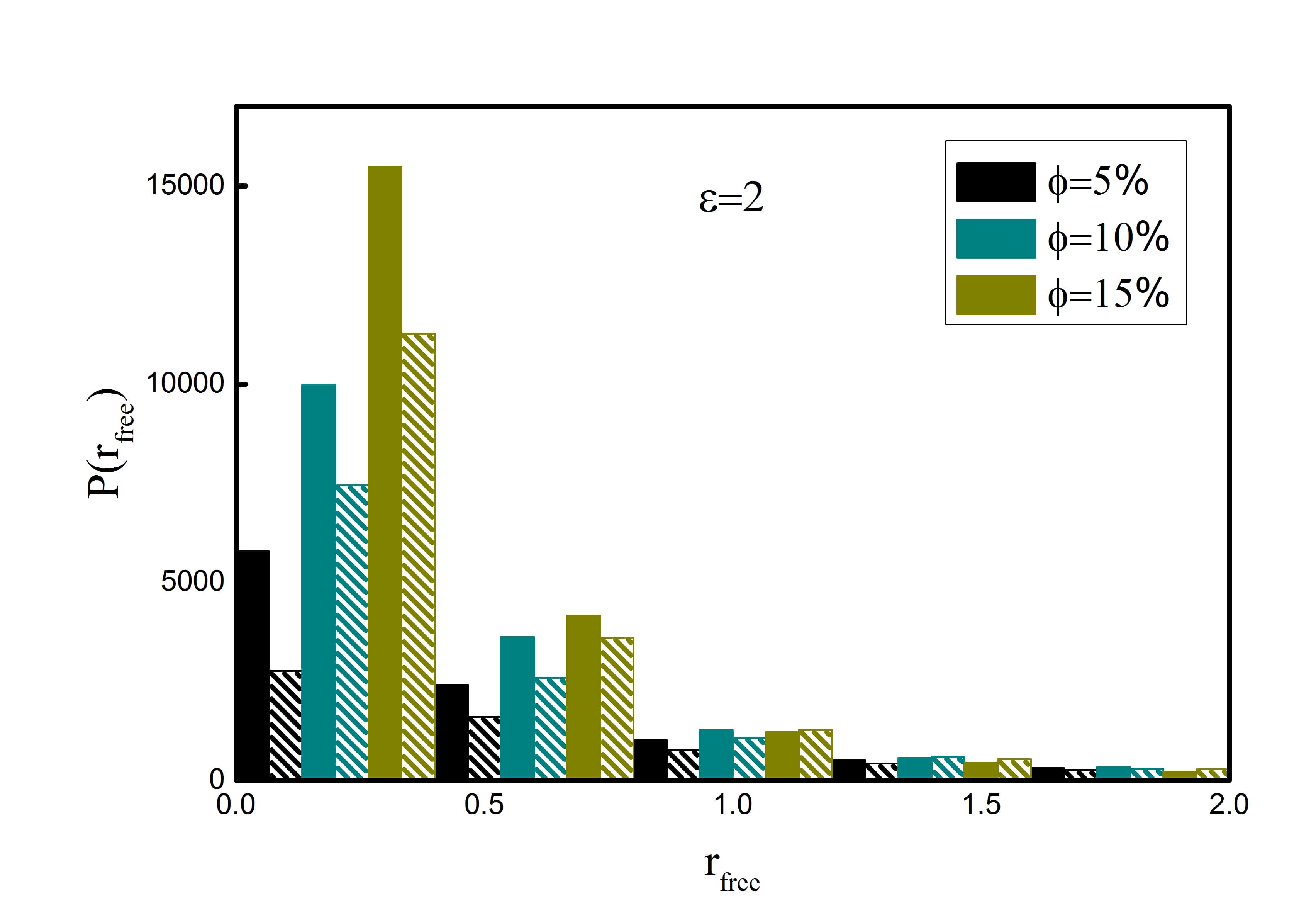} (a)
     \includegraphics[width=.5\textwidth]{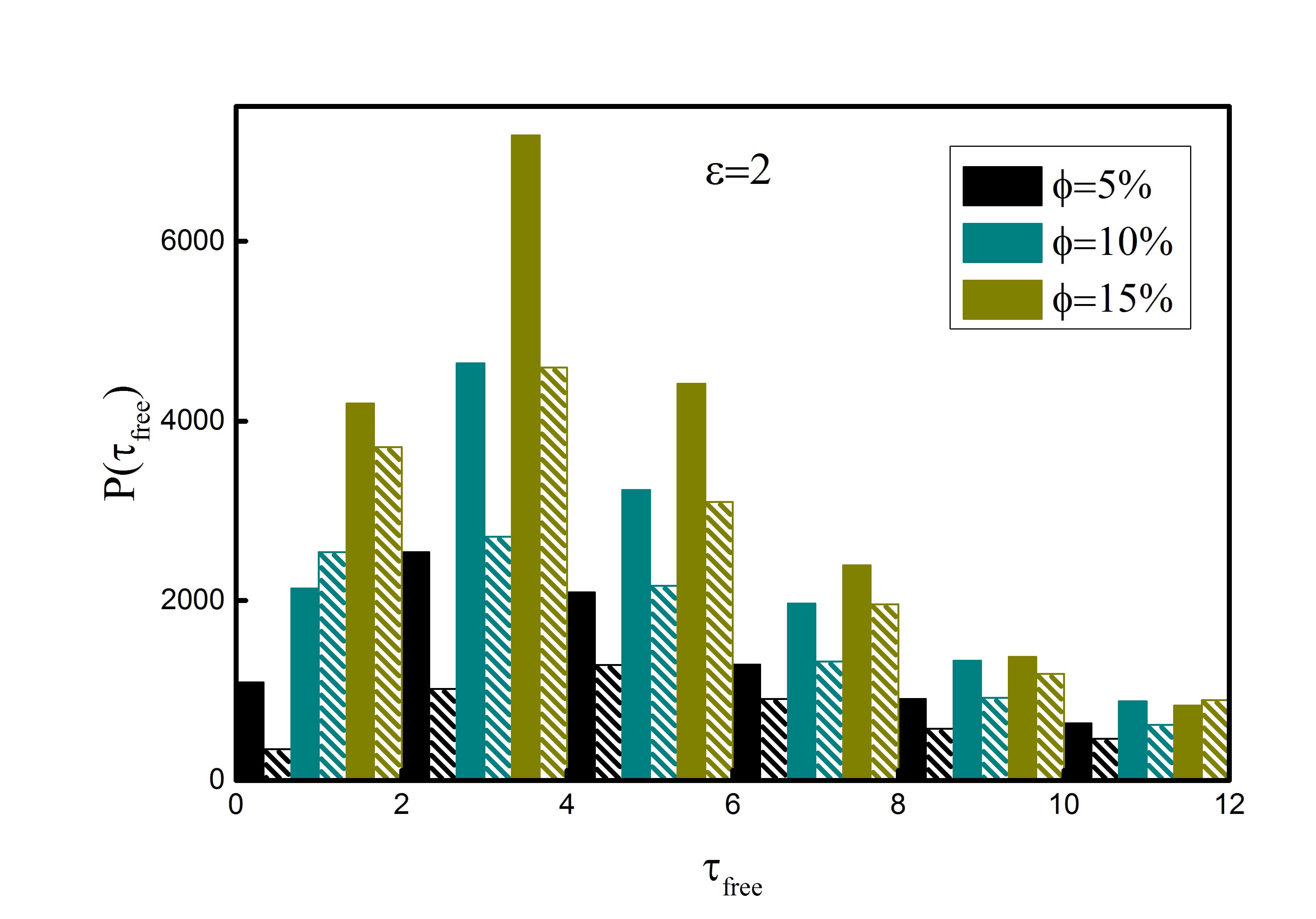} (b)
\end{tabular}
   \caption{Histograms of (a) the distance travelled by the tracer in the unbound state (b) the time spent in the unbound state at different degrees of volume occupancy. The solid bars correspond to the tracer in the presence of mobile polymers and the bars filled with dashed lines correspond to the tracer in the presence of frozen polymers.}
   \label{fig:8}
 \end{figure}

\begin{figure}
 \centering
   \begin{tabular}{@{}cccc@{}}
     \includegraphics[width=.5\textwidth]{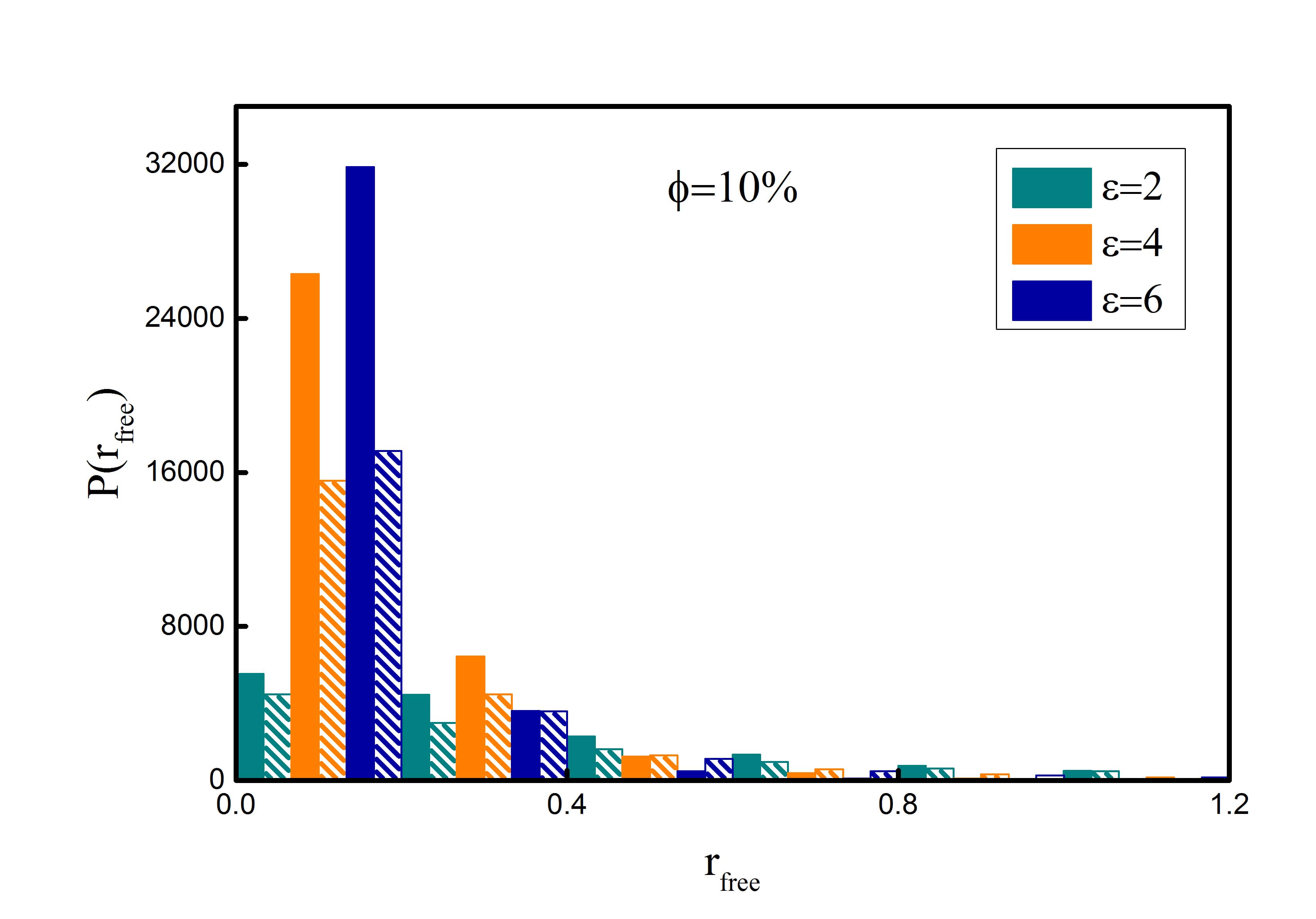} (a)
     \includegraphics[width=.5\textwidth]{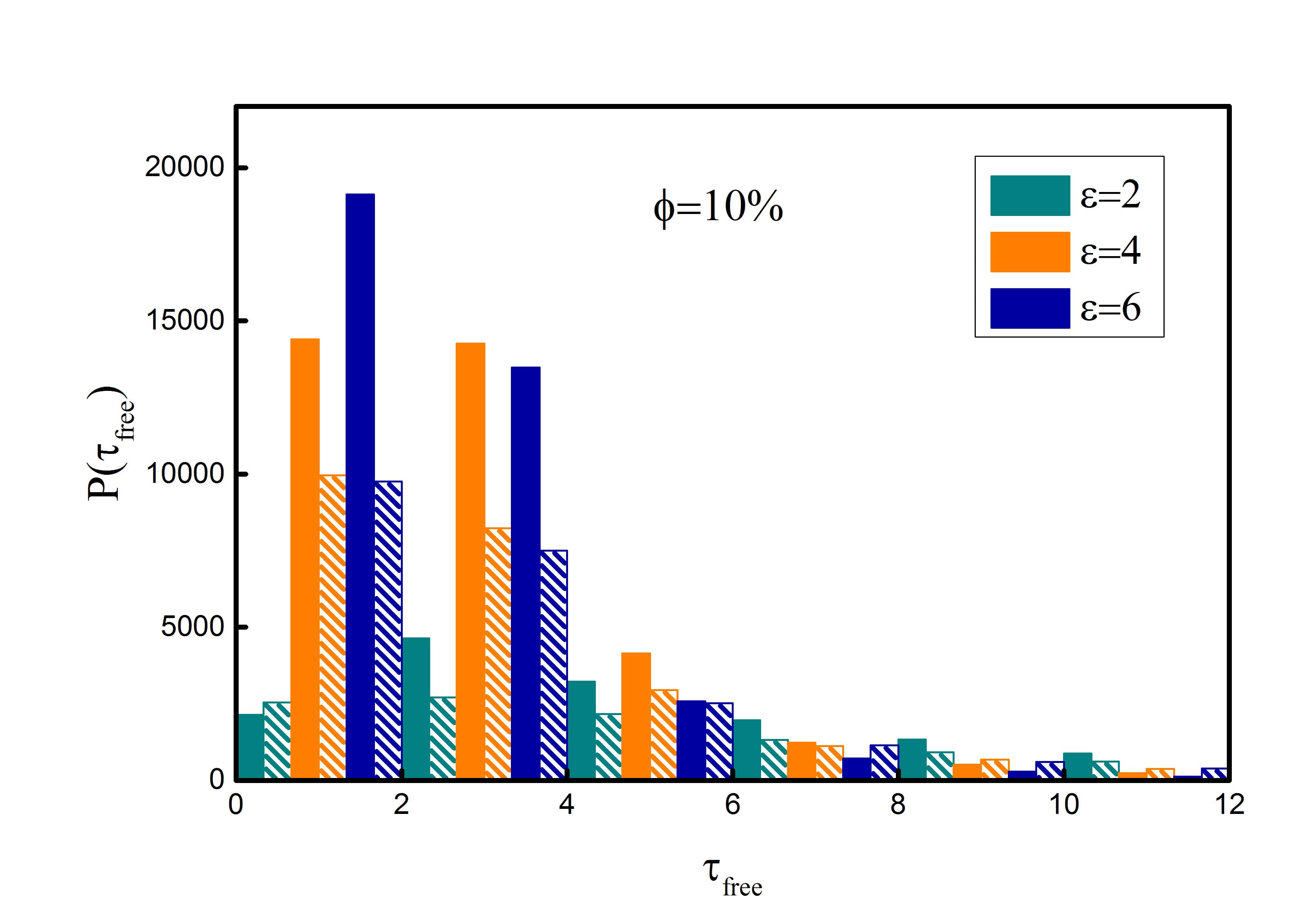} (b)
\end{tabular}
   \caption{Histograms of (a) the distance travelled by the tracer in the unbound state (b) the time spent in the unbound state at different levels of tracer binding zone interaction. The solid bars correspond to the tracer in the presence of mobile polymers and the bars filled with dashed lines correspond to the tracer in the presence of frozen polymers.}
   \label{fig:9}
 \end{figure}

In this section we present the the statistics of the binding and unbinding processes of the tracer in the trapping zones of the polymers. There is no unique way of defining the trapping. In our case, the tracer is regarded to be trapped when the tracer is within a minimum distance from any two or more binding monomers of any of the polymer present in the simulation box. The minimum distance is
less than or equal to $1.1\sigma$, where $\sigma=r_{tracer}+r_{monomer}$, around the Lennard-Jones minima, otherwise it is considered to be free or unbound. It is very evident from here that the statistics obtained from this representation of trapping
will vary if a different definition of trapping is followed, however it is expected that the overall trends will always remain the same. However there could be a situation that the tracer is caged but in our definition it is not trapped, especially when the cages are big ($\sim 2 $ in length scale). We calculate the distance between the tracer with every binding monomers
 in the system at each time step and even if the binding monomers change in two consecutive steps, the tracer is considered to be trapped.
Fig. (\ref{fig:6}a) shows how the distribution of the distance travelled by the tracer in the trapped state vary with changing the volume fraction
and Fig. (\ref{fig:6}b) shows the distribution of the time spent by the tracer in the trapped state.
The histogram plots show the probability of the tracer to be trapped for short time is the most likely event and it then decays with the increasing trapping time. It should also be mentioned that the inset of Fig. (\ref{fig:6}b) represents the log-log plot of the distribution of the trapping time and it clearly shows that the distribution does not follow a simple power law. On increasing the volume fraction the trapping probability
increases. This is because, as the number of polymers increases, there are more number of polymer traps available for the tracer to bind with. Therefore when the tracer spends longer time in the trapped state it can
even travel longer distances, but this could be in a cage or outside a cage. One should notice that the decay rate of the histogram peaks are much slower in case of frozen traps in comparison to the mobile ones.
 This indicates that the tracer spends more time in trapped
state when the polymers are immobile, while in presence of mobile traps the probability of staying trapped for long time is less. Fig.(\ref{fig:7}) shows
 the similar distribution function of $r_{trap}$ and $\tau_{trap}$ for different values of $\epsilon$. On increasing the binding strength the
trapping probability increases which gets reflected in the distribution plots. In this case too, the probability of spending longer time in the trapped
state is higher in the presence of frozen polymers. From the corresponding average values given in  Table (\ref{tablephi}) and Table (\ref{tableeps}) it can be seen
the $<r_{trap}>$ and $<\tau_{trap}>$ are always higher in case of frozen polymers and with increasing $\phi$ and $\epsilon$ the average values increases. Although there is slight
discrepancy in the average values for $\epsilon=6$ (not shown) in case of frozen polymers, where the average values decreases in comparison to $\epsilon=4$. As already mentioned in this particular case the tracer practically remains static, the dynamics becomes non-ergodic and we do not have enough statistics to calculate averages.
The distributions for the distance ($r_{free}$)
and time ($\tau_{free}$) covered in unbound state are shown in  Fig.(\ref{fig:8}) and Fig.(\ref{fig:9}) and corresponding average values are given in  Table (\ref{tablephi}) and Table (\ref{tableeps}).

\begin{table}[tbp]
\begin{tabular}{|c||c|c|c|c|c|c|c|c|}
\hline
Average values & \multicolumn{2}{c|}{$<r_{trap}>$ } & \multicolumn{2}{c|}{$<\tau_{trap}>$}& \multicolumn{2}{c|}{$<r_{free}>$} & \multicolumn{2}{c|}{$<\tau_{free}>$}\\ \hline
  Volume fraction&Mobile& Frozen & Mobile& Frozen & Mobile & Frozen & Mobile& Frozen \\ \hline
$\phi=5\%$&0.11&0.08&1.22&1.38&2.37&5.08&11.87&21.25 \\ \hline
$\phi=10\%$&0.13&0.14&1.3&1.84&1.49&1.56&7.48&10.96 \\ \hline
$\phi=15\%$&0.13&0.16&1.39&1.81&1.03&1.88&5.27&7.48 \\ \hline
 \end{tabular}
\caption{The table shows the average distances travelled and the average times spent by the tracer in the trapped as well as in the free states at different monomer volume fraction.}
\label{tablephi}
\end{table}

 \begin{table}[tbp]
 \begin{tabular}{|c||c|c|c|c|c|c|c|c|}
 \hline
 Average values &\multicolumn{2}{c|}{$<r_{trap}>$} & \multicolumn{2}{c|}{$<\tau_{trap}>$}& \multicolumn{2}{c|}{$<r_{free}>$} & \multicolumn{2}{c|}{$<\tau_{free}>$}\\ \hline
   Binding affinity &Mobile& Frozen & Mobile& Frozen & Mobile & Frozen & Mobile & Frozen  \\ \hline
$\epsilon=2$& 0.13&0.14&1.3&1.84&1.49&1.56&7.48&10.96\\ \hline
$\epsilon=4$&0.2&3.82&1.84&55.27&0.42&31.53&2.45&4.14 \\ \hline
\end{tabular}
 \caption{The table shows the average distances travelled and the average times spent by the tracer while it is trapped as well while it is free at different values of the binding strengths of the traps.}
 \label{tableeps}
 \end{table}

\subsection{Control simulations}

\textbf{Polymers without binding zones:} To investigate only the effect of crowding in absence of any binding in tracer trapping, another set of simulations are performed with a volume fraction of $\phi=15\%$ consisting of polymers with no binding zones. This
means all the polymers in the system have only repulsive (WCA) interaction with the tracer \cite{andersen}. Keeping the size of tracer and the monomers same, we see almost no trapping. The tracer remains free most of the time and in the absence of attractive interaction even the higher population of the polymers does not lead to trapping.

\textbf{Polymers with all attractive monomers:} To study the effect of only binding affinity we carry out another set of control simulations where all the monomers in the polymer chains have attractive interaction with the tracer, therefore the whole polymer acts as a trap for the tracer. To minimize the effect of number of the traps we keep only $10$ polymers in the system each having $20$ monomers. The binding affinity between the tracer and the monomers are quite high, $\epsilon=4$ and all the particles are of same size. From these set of simulations we see even when the number of traps is less, the tracer is trapped for a considerable length of time and shows subdiffusive behavior with an exponent of $\beta\sim0.45$. Although if compared with the case at $\phi=10\%$ and $\epsilon=4$, where the tracer shows slightly more subdiffusive behavior and the average trapping time is also higher. Thus it can be confirmed that in this semi-dilute regime, it is the binding affinity of the traps that plays prevalent role in trapping rather than the the number of traps.

\subsection{Effect of Tracer size}

   \begin{figure}
 \centering
  \includegraphics[width=.4\textwidth]{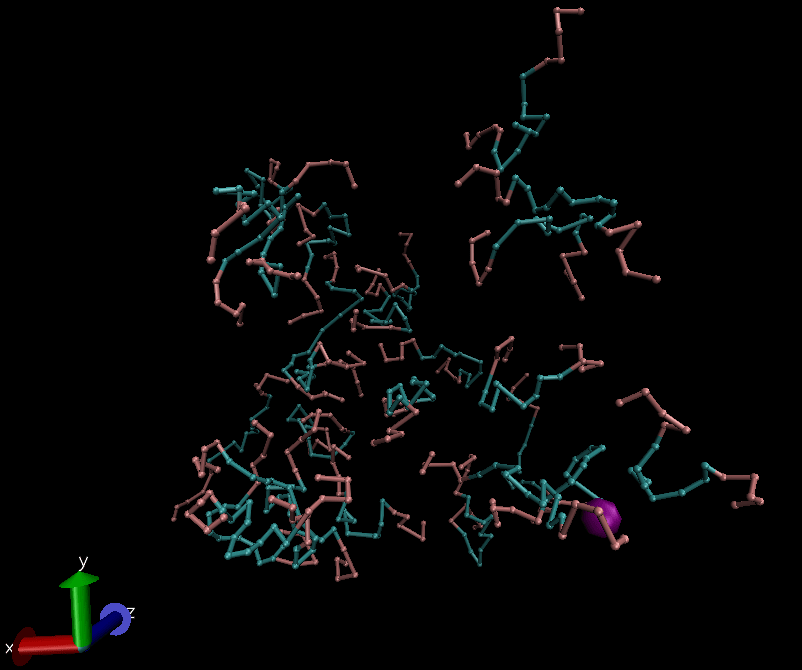}
      \caption A typical snap shot of the bigger tracer and the polymers. The tracer is shown in purple and the binding zones of the polymers are shown in cyan.
   \label{fig:1}
 \end{figure}

\begin{figure}
  \centering
    \begin{tabular}{@{}cccc@{}}
      \includegraphics[width=.5\textwidth]{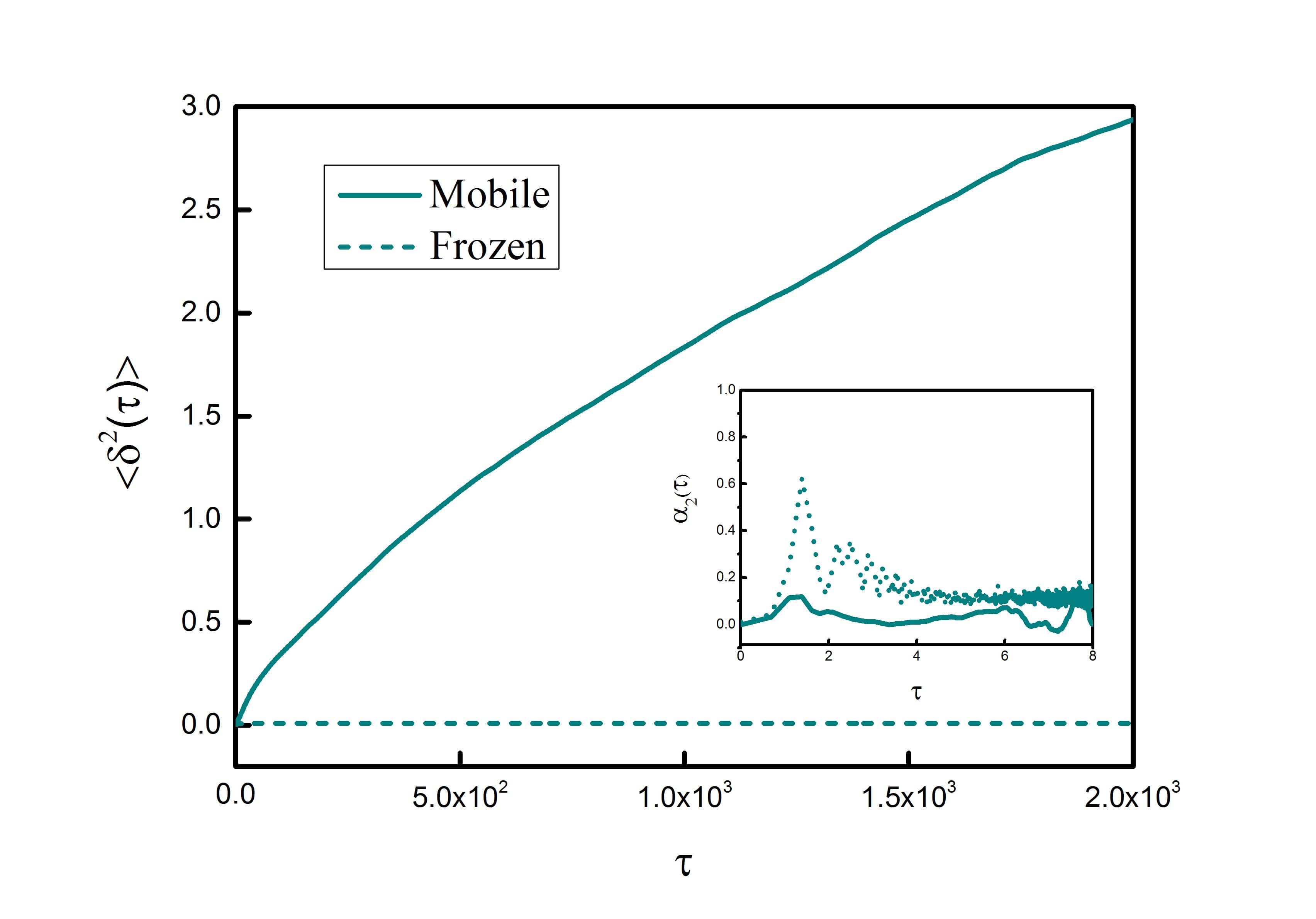} (a)
      \includegraphics[width=.5\textwidth]{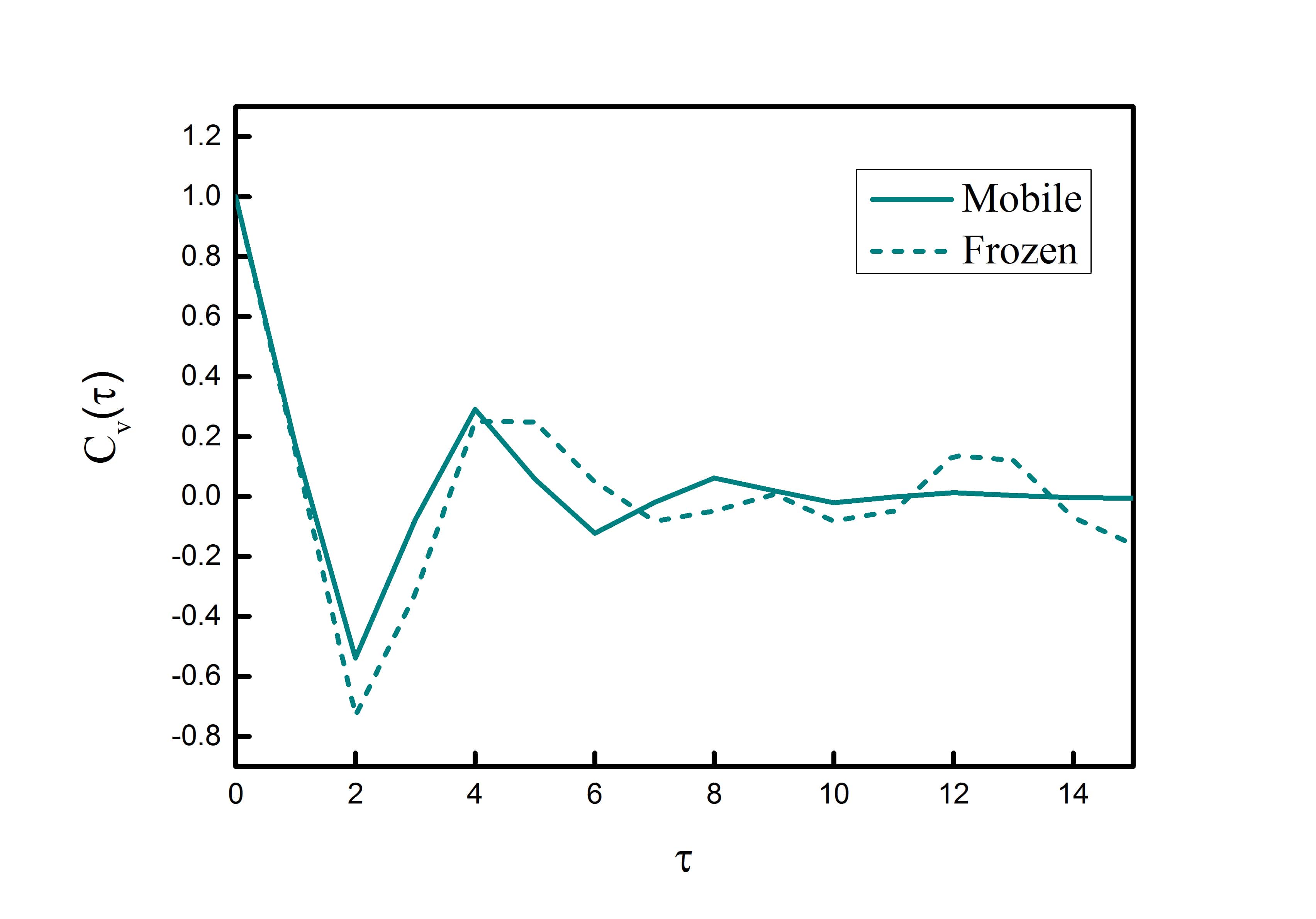} (b)
 \end{tabular}
    \caption{(a) Plot of $\left<\overline{\delta^{2}(\tau)}\right>$ vs $\tau$ . The inset shows the Log-linear plot of non-Gaussianity parameter ($\alpha_2(\tau)$) vs $\tau$ for the same (b) Log-linear plot of the velocity autocorrelation function ($C_v(\tau)$) vs time ($\tau$). The values of the parameters chosen for the bigger tracer are $\phi=10\%$ and $\epsilon=2$.  }
    \label{fig:10}
  \end{figure}

To study the effect of the size of the tracer in binding and unbinding processes we perform another set of simulations with a tracer, five times bigger than the previous one keeping all the other parameters unchanged. The tracer in this case have a radius of $2.5$ although the size of monomers in the polymers remain the same ($0.5$). A VMD \cite{schultenvmd} snapshot of the simulation can be seen in Fig.(\ref{fig:1}b). With the bigger tracer the simulations are performed at volume fraction, $\phi=10\%$, and the binding affinity of the trapping zones are fixed at $\epsilon=2$. As the size of the tracer is increased, lesser number of polymers are included in the simulation  to maintain the volume fraction $\phi\sim10\%$ . Simulations are done in the presence of mobile and frozen polymers
 and ten trajectories are generated for each case.
Fig.(\ref{fig:10}a) is the plot of $\left<\overline{\delta^{2}(\tau)}\right>$ vs time and it can be seen that the $\left<\overline{\delta^{2}(\tau)}\right>$ for the bigger tracer even in the presence of mobile polymers is very low and it becomes almost negligible in
the presence of frozen polymers. In the inset of Fig.(\ref{fig:10}a) we show the non-Gaussianity parameters which show the diffusion to be almost Gaussian in the presence of mobile
polymers and weakly non-Gaussian in the presence of static polymers. Fig.(\ref{fig:10}b) shows the velocity autocorrelation functions for the two different cases and both of them have negative values at short time.  From the calculation of trapping time (not shown here) with the same conditions, where the tracer is considered to be trapped when that is within a distance of
$1.1\sigma$, where $\sigma=r_{tracer}+r_{monomer}=1.5$. The tracer is found to trapped in the entire simulation time even in the presence of mobile polymers. The size of the
tracer is found to play a crucial role in binding-unbinding process and a bigger tracer always facilitate trapping.
Our observations show similar trend as found in recent experiments on the tracer diffusion in polymer gels with tracers of varying sizes \cite{granickmacmol, hujpclet}. For example, as found in recent experiment on the tracer dynamics in thermoreversible gels \cite{granickmacmol}, in our case too, the bigger tracers show subdiffusive behavior whereas the smaller tracers exhibit normal diffusion unless the background is very sticky or frozen. On the other hand, in an another experiment the dynamics of bigger tracers in polymer gel is found  out to be Gaussian \cite{hujpclet} and this is exactly what we find in our simulations.

 \section{Conclusions}

Motivated by recent experiments on the tracer diffusion in polymeric materials \cite{granickpnas,granickacsnano,hujpclet}, we investigate the tracer diffusion in a polymer solution by molecular dynamics simulations. The polymers in our model have specific
binding zones to trap the tracer and since many in numbers can form cages either transient or permanent depending on whether these polymers are mobile or frozen. Our simulations confirm that it is rather the higher binding strength than the extent of crowding that makes the tracer diffusion subdiffusive. With frozen polymers the tracer exhibits jiggling motion in a cage, followed by cage to cage jumps resembling CTRW and resulting a non-Gaussian statistics but whether diffusive or subdiffusive that depends on the volume fraction and the binding strength of the traps. However, when the polymers are mobile, subdiffusion is observed only when the volume fraction or the binding strengths are high. We also find that with increasing binding strength and the population of the polymers, the probability of the tracer being trapped increases. However,
the number of traps hardly facilitate trapping, since in the absence of any attractive interaction between the tracer and the polymers, the tracer rarely gets trapped, whereas in the presence of a small number of polymer traps the tracer shows trapping if the binding affinity of the traps is higher. Therefore it is the binding affinity rather than the number of traps that facilitates trapping. The system remains in a semi-dilute regime even at the maximum volume fraction we considered. But in future we would like to explore a more crowded environment relevant in the context of biological cells \cite{Elcock}. Another interesting observation is that the trapping probability increases with the increasing size of the tracer and the dynamics is still weakly non-Gaussian unless the environment is mobile. We hope that our study will help in understanding tracer diffusion in crowded environment and shed light on how differently mobile and the static crowders control the process.

\section{Acknowledgements}
NS thanks Surya K. Ghosh for many valuable discussions. RC thanks IRCC IIT Bombay (Project Code: 12IRCCSG046) and DST (Project No. SB/SI/PC-55/2013) for fundings.

\bibliographystyle{apsrev}


\end{document}